\documentclass[onecolumn]{aa}

\usepackage[T1]{fontenc}
\usepackage{pslatex}
\usepackage{amsmath}
\usepackage{amsfonts}
\usepackage{xcolor}
\usepackage{url}
\usepackage{bm}
\usepackage{graphicx}
\usepackage{amssymb}
\usepackage{soul}
\usepackage{booktabs}
\usepackage{array}
\usepackage[utf8]{inputenc}
\inputencoding{latin1}
\inputencoding{utf8}
\usepackage{appendix}

\def\lsim{ \lower .75ex\hbox{$\sim$} \llap{\raise .27ex \hbox{$<$}} }
\def\gsim{ \lower .75ex \hbox{$\sim$} \llap{\raise .27ex \hbox{$>$}} }

\newcommand{\bi}{\begin{itemize}}
\newcommand{\ei}{\end{itemize}}

\newcommand{\pa}{\partial}

\usepackage[colorlinks=true,linkcolor=blue,citecolor=blue]{hyperref}

\title{Filamentation of the electromagnetic precursor \\ in relativistic quasi-perpendicular electron-positron shocks}
\titlerunning{Electromagnetic precursor in relativistic shocks}
\authorrunning{Sobacchi et al.}

\author{
E. Sobacchi\inst{1,2}
\and
Y. Lyubarsky\inst{3}
\and
L. Sironi\inst{4,5}
\and
M. Iwamoto\inst{6}
}

\institute{
Gran Sasso Science Institute, viale F.~Crispi 7, L’Aquila 67100, Italy\\
\email{emanuele.sobacchi@gssi.it}
\and
INFN -- Laboratori Nazionali del Gran Sasso, via G.~Acitelli 22, Assergi 67100, Italy
\and
Physics Department, Ben-Gurion University, Be'er-Sheva 84105, Israel
\and
Department of Astronomy and Columbia Astrophysics Laboratory, Columbia University, 550 W 120th St, New York, NY 10027, USA
\and
Center for Computational Astrophysics, Flatiron Institute, 162 5th Avenue, New York, NY 10010, USA
\and
Graduate School of System Informatics, Kobe University, 1-1 Rokkodai-cho, Nada, Kobe, Hyogo 657-8501, Japan
}
\date{}

\begin{document}

\abstract{
We present a scenario that could explain non-thermal particle acceleration in relativistic quasi-perpendicular electron-positron shocks, such as the termination shock of pulsar wind nebulae. The shock produces a strong electromagnetic precursor that propagates into the upstream plasma, which is initially threaded by a uniform background magnetic field. We show that the filamentation instability breaks the precursor into radiation filaments parallel to the shock normal. The transverse scale of the filaments is of the order of a few plasma skin depths. In the shock frame, the bulk Lorentz factor of the upstream plasma is significantly reduced inside the radiation filaments. Then, the instability produces a relativistic shear flow with strong velocity gradients on kinetic scales. The velocity gradients distort the background magnetic field lines, and generate a magnetic field component parallel to the shock normal that reverses across each radiation filament, a configuration that could trigger magnetic reconnection in the upstream plasma. These effects may accelerate particles before the plasma enters the shock.
}


\maketitle
\boldsymbol{}

\section{Introduction}

Theoretical models of relativistic quasi-perpendicular electron-positron shocks suggest that diffusive shock acceleration is inhibited because particles sliding along the magnetic field lines cannot cross the shock multiple times. Then, the downstream particles should have a quasi-thermal distribution \citep[][]{BegelmanKirk1990, Gallant+1992, SironiSpitkovsky2009}. Pulsar wind termination shocks are prototypical examples of such shocks. Multiwavelength observations show that the pulsar wind nebulae are filled with non-thermal particles \citep[][]{GaenslerSlane2006, Hester2008, BuhlerBlandford2014, Kargaltsev+2015, Reynolds+2017}. The hardness of the radio spectrum implies that the acceleration mechanism transfers most of the available energy to a small fraction of the particles, which is not what one would expect in diffusive shock acceleration. However, the morphology of the nebulae indicates that non-thermal particles are accelerated at the termination shock \citep[][]{KomissarovLyubarsky2003, KomissarovLyubarsky2004, DelZanna+2004, DelZanna+2006, Porth+2014, Olmi+2015}. It is clear that theoretical models of shocks lack some key element that allows the acceleration of non-thermal particles.

Most studies attempt to revive particle acceleration by modifying the properties of the upstream plasma. Some possibilities are as follows. (i) The plasma carries alternating magnetic field lines separated by current sheets, which reconnect at the termination shock \citep[][]{Lyubarsky2003, PetriLyubarsky2007, LyubarskyLiverts2008, SironiSpitkovsky2011}. However, reconnection-driven acceleration of non-thermal particles requires extremely large particle multiplicities. (ii) The plasma is contaminated with protons \citep[][]{Hoshino+1992, AmatoArons2006}. However, in order to efficiently accelerate the electron-positron pairs, protons should carry a significant fraction of the energy flux, which seems inconsistent with the modeling of pulsar wind nebulae \citep[][]{Bucciantini+2011}. (iii) The magnetic field intensity has a large scale gradient in the latitudinal direction \citep[][]{CeruttiGiacinti2020}. In this scenario, particles are accelerated by shear flow in an elongated cavity that forms in the equatorial region of the shock and by turbulence that develops in the shock downstream.

Another line of research emphasizes the role of the electromagnetic precursor in the shock dynamics. It is now well established that relativistic quasi-perpendicular electron-positron shocks produce strong electromagnetic waves that propagate into the upstream plasma \citep[][]{Langdon+1988, Gallant+1992, Iwamoto+2017, Iwamoto+2018, PlotnikovSironi2019, BabulSironi2020, Margalit+2020, Sironi+2021}. The strength parameter of the electromagnetic precursor can be large, namely $a_0>1$ (the strength parameter, $a_0$, is defined as the peak transverse component of the particles four-velocity in units of the speed of light). \citet{Lyubarsky2018, Lyubarsky2019} argues that absorption of the strong electromagnetic precursor could accelerate particles. These studies consider synchrotron absorption and induced Compton scattering as absorption mechanisms. However, the applicability of these absorption mechanisms to the termination shock of pulsar wind nebulae is not discussed in detail.

The filamentation of the electromagnetic precursor has received limited attention so far, although \citet{Lyubarsky2018, Lyubarsky2019} correctly argues that filamentation can be important for shock dynamics. The main difficulty is that one should consider the regime of large wave strength parameters, $a_0>1$. The filamentation instability has been extensively studied in different fields, including laser-plasma interaction \citep[][]{Kaw+1973, Drake+1974, Max+1974} and fast radio bursts \citep[][]{Sobacchi+2021, Sobacchi+2022, Sobacchi+2023, Ghosh+2022, Iwamoto+2023}. These studies focus on the non-relativistic limit $a_0\ll 1$. In this limit, the physics of the instability is understood as follows. Modulations of the radiation intensity in the direction transverse to the wave propagation are expected to be unstable because the ponderomotive force pushes particles out of regions where the intensity is enhanced. The resulting modulation of the plasma refractive index creates a converging lens that further enhances the radiation intensity, thus forming a positive feedback loop. In our previous work \citep{Sobacchi+2024a}, we studied the filamentation of electromagnetic waves in unmagnetized electron-positron plasmas in the regime $a_0>1$. However, our results cannot be applied to the precursor of relativistic shocks, where the upstream plasma is magnetized.

In this paper, we study the filamentation instability in magnetized electron-positron plasmas. We focus on the regime $a_0>1$, as appropriate for relativistic shocks. We show that the electromagnetic precursor at the termination shock of pulsar wind nebulae is almost certainly filamented. The instability has important implications for shock dynamics. In the shock frame, the bulk Lorentz factor of the upstream plasma is significantly reduced when the plasma is illuminated by the electromagnetic precursor. Then, the flow becomes sheared when the precursor is filamented. The shear flow distorts the background magnetic field lines, and generates a field component parallel to the shock normal that reverses across each radiation filament, potentially triggering magnetic reconnection. Relativistic shear flow and magnetic reconnection may accelerate particles before the plasma enters the shock.

We show that the physics of the filamentation instability can be modified by the presence of a strong background magnetic field. As we explained above, the instability develops because the density of the plasma increases outside radiation filaments, and decreases inside the filaments. When the magnetization of the plasma is large, the ponderomotive force is not the main reason why the density is modified. Instead, the dominant effect is the following. Outside radiation filaments the density increases because the tension of the distorted magnetic field lines pushes the plasma in the direction of wave propagation, whereas inside the filaments the density decreases because the direction of the tension force is reversed.

The paper is organized as follows. In Sect.~\ref{sec:precursor}, we review the properties of the electromagnetic precursor. In Sect.~\ref{sec:equations}, we derive the equations that govern the evolution of the precursor. In Sect.~\ref{sec:filamentation}, we calculate the growth rate and the most unstable wave number of the filamentation instability. In Sect.~\ref{sec:shock}, we discuss the implications of the instability for relativistic shocks. In Sect.~\ref{sec:concl}, we summarize our conclusions. Readers not interested in technical details may skip Sects.~\ref{sec:equations} and \ref{sec:filamentation}.

\section{Properties of the electromagnetic precursor}
\label{sec:precursor}

The properties of the electromagnetic precursor are extensively discussed in the literature \citep[][]{Langdon+1988, Gallant+1992, Iwamoto+2017, Iwamoto+2018, PlotnikovSironi2019, BabulSironi2020, Margalit+2020, Sironi+2021}. In the following, we summarize the available estimates for the frequency and strength parameter of the precursor. 

In the pulsar frame, the termination shock is stationary. The pulsar wind has a large Lorentz factor, $\Gamma_{\rm w}$, with respect to the shock. In the proper frame of the pulsar wind (hereafter, the ``wind frame''), the particles ahead of the precursor are at rest. We define the plasma frequency as $\omega_{\rm P}=\sqrt{8\pi n_0 e^2/m}$, where $n_0$ is the proper positron number density ahead of the precursor, and $e$ and $m$, respectively, are the charge and the mass of the positron. We define the Larmor frequency as $\omega_{\rm L}=eB_{\rm bg}/mc$, where $B_{\rm bg}$ is the background magnetic field in the wind frame ahead of the precursor, and $c$ is the speed of light. The bulk Lorentz factor of the downstream plasma with respect to the shock is of the order of $\sqrt{1+\sigma}$, where the magnetization is defined as $\sigma=\omega_{\rm L}^2/\omega_{\rm P}^2$.

In the wind frame, the frequency of the precursor is \citep[][]{Iwamoto+2017, PlotnikovSironi2019, Sironi+2021}
\begin{equation}
\label{eq:omegaprec}
\omega_0 \simeq 3\;\Gamma_{\rm w}\omega_{\rm P} = 3\;\frac{\Gamma_{\rm w}}{\sqrt{\sigma}}\;\omega_{\rm L} \;.
\end{equation}
When $\sigma>1$, the strength parameter can be presented as $a_0\simeq 0.03\;\Gamma_{\rm w,d}$, where $\Gamma_{\rm w,d}=\Gamma_{\rm w}/\sqrt{1+\sigma}$ is the Lorentz factor of the wind with respect to the downstream plasma \citep[][]{PlotnikovSironi2019, Sironi+2021}. For a given $\Gamma_{\rm w,d}$, the strength parameter increases for decreasing magnetization when $0.1<\sigma<1$, and then decreases when $10^{-3}<\sigma<0.1$ (see Fig.~11 of \citealt[][]{Iwamoto+2017}, and Fig.~8 of \citealt[][]{PlotnikovSironi2019}). In this paper, we assume
\begin{equation}
\label{eq:aprec}
a_0 \simeq 0.03\; \frac{\Gamma_{\rm w}}{\sqrt{1+\sigma}} \;,
\end{equation}
which is the exact expression when $\sigma>1$, and is accurate within a factor of a few when $10^{-3}<\sigma<1$. Eqs.~\eqref{eq:omegaprec} and \eqref{eq:aprec} are also valid in the regime $a_0>1$ \citep{Margalit+2020}. In the wind frame, the magnetic field of the precursor, $B_0=a_0 mc\omega_0/e$, can be much larger than the background magnetic field. Using Eqs.~\eqref{eq:omegaprec} and \eqref{eq:aprec}, one finds $B_0/B_{\rm bg}=a_0\omega_0/\omega_{\rm L}\simeq 0.1\;\Gamma_{\rm w}^2/\sqrt{\sigma (1+\sigma)}$.

In the following, we estimate the frequency and strength parameter of the electromagnetic precursor at the termination shock of the Crab pulsar wind. The Crab pulsar has a period $P=33{\rm\; ms}$, and a magnetic moment $\mu=5\times 10^{30}{\rm\; G\; cm^3}$. At the light cylinder radius, $R_{\rm LC}=cP/2\pi$, the magnetic field in the pulsar frame is $B_{\rm LC}=\mu/R_{\rm LC}^3$, and the particle number density is $N_{\rm LC}=\kappa N_{\rm GJ}$, where $N_{\rm GJ}=B_{\rm LC}/ecP$ is the Goldreich-Julian density \citep[][]{GoldreichJulian1969}, and $\kappa$ is the particle multiplicity. Conservation of energy implies $\kappa\Gamma_{\rm w}(1+\sigma) = eB_{\rm LC}P/4\pi mc$ \citep[][]{Kargaltsev+2015}. The latter relation can be presented as $\Gamma_{\rm w}=6\times10^5 (1+\sigma)^{-1} \kappa_5^{-1}$, where $\kappa_5=\kappa/10^5$. Then, Eq.~\eqref{eq:aprec} implies $a_0\simeq 2\times 10^4 (1+\sigma)^{-3/2} \kappa_5^{-1}$.

The particle multiplicity, $\kappa$, and the magnetization at the termination shock, $\sigma$, are difficult to calculate from first principles. The modeling of the Crab pulsar wind nebula suggests $\kappa> 10^5$ \citep[][]{Bucciantini+2011}, which is consistent with the multiplicity that one would expect from pair production in the pulsar polar cap \citep[][]{TimokhinHarding2015, TimokhinHarding2019}. The magnetization of the upstream plasma at the termination shock, $\sigma$, depends on the amount of magnetic energy that is dissipated by reconnection in the wind \citep[][]{LyubarskyKirk2001, KirkSk2003}. Recent fully kinetic simulations show that alternating magnetic field lines could reconnect relatively close to the light cylinder \citep[][]{CeruttiPhilippov2020}. After reconnection occurs, the plasma is threaded by a uniform magnetic field. In the equatorial region, where stripes of alternating magnetic field are roughly symmetric before reconnection occurs, one would expect $\sigma<1$. In any case, the wave strength parameter is large for any reasonable value of $\kappa$ and $\sigma$, since condition $a_0>1$ requires $\kappa< 2\times 10^9 (1+\sigma)^{-3/2}$.

The Crab pulsar wind terminates at the radius $R_{\rm TS}=4\times 10^{17} {\rm\; cm}$ \citep[][]{Weisskopf+2000}. At the termination shock, the proper number density of the positrons is $n_0=N_{\rm LC}R_{\rm LC}^2/2\Gamma_{\rm w}R_{\rm TS}^2$. Taking into account that $N_{\rm LC}=\kappa B_{\rm LC}/ecP= 3\times 10^{11} \kappa_5 {\rm\; cm}^{-3}$, one can calculate the plasma frequency, which is $\omega_{\rm P}= 0.02\;\kappa_5\sqrt{1+\sigma}{\rm\; Hz}$. Then, from Eq.~\eqref{eq:omegaprec} one finds $\omega_0\simeq 30{\rm\; kHz}/\sqrt{1+\sigma}$.

\section{Evolution of the electromagnetic precursor}
\label{sec:equations}

We consider a linearly polarized quasi-monochromatic wave packet that propagates in a cold magnetized electron-positron plasma. In the wind frame, the wave frequency is $\omega_0$, and the wave vector is $k_0 {\bm e}_z$. The electric field of the wave is directed along ${\bm e}_y$, and the background magnetic field is directed along ${\bm e}_x$. The wave polarization corresponds to the extraordinary mode, which is the dominant component of the electromagnetic precursor for large magnetization \citep[][]{Sironi+2021}.

We work in the frame that moves with the group velocity, $k_0/\omega_0$, along ${\bm e}_z$ (hereafter, the ``wave frame''). In our units, the speed of light is $c=1$.  As discussed by \citet{Clemmow1974}, in this frame the wave vector vanishes, and the wave frequency is
\begin{equation}
\omega= \sqrt{\omega_0^2 - k_0^2}\;.
\end{equation}
In the wave frame, particles ahead of the wave packet have a large four velocity, ${\bm u}_0=-u_0{\bm e}_z$, where $u_0=k_0/\omega$ (the corresponding Lorentz factor is $\gamma_0=\omega_0/\omega$). The magnetic field is ${\bm B}=(m/e)\gamma_0\omega_{\rm L}{\bm e}_x$. Since in the wave frame the particles ahead of the wave packet are moving, there is an electric field ${\bm E}=(m/e)u_0\omega_{\rm L}{\bm e}_y$. Inside the wave packet, the magnetic field can be presented as ${\bm B}=(m/e)[\gamma_0\omega_{\rm L}{\bm e}_x+\nabla\times \tilde{\bm a}]$, and the electric field can be presented as ${\bm E}=(m/e)[u_0\omega_{\rm L}{\bm e}_y-\pa\tilde{\bm a}/\pa t]$, where $(m/e)\tilde{\bm a}$ is the vector potential of the wave. We work in the Coulomb gauge, which is $\nabla\cdot\tilde{\bm a}=0$. As we demonstrate in Appendix \ref{sec:charge}, the electrostatic potential vanishes because in electron-positron plasmas the wave does not produce any charge separation. As discussed by \citet{BourdierFortin1979}, the equation of motion of the particles can be presented as
\begin{equation}
\label{eq:euler}
\frac{\pa}{\pa t}\left({\bm u}_\pm\pm\tilde{\bm a}\right) -\frac{{\bm u}_\pm}{\gamma_\pm} \times \left[\nabla\times\left({\bm u}_\pm \pm \tilde{\bm a} \right)\right] = -\nabla\gamma_\pm \pm\left(\frac{{\bm u}_\pm}{\gamma_\pm}+\frac{u_0}{\gamma_0}{\bm e}_z\right)\times\gamma_0\omega_{\rm L}{\bm e}_x \;,
\end{equation}
where ${\bm u}_\pm$ is the four-velocity, and $\gamma_\pm$ is the Lorentz factor (the plus and minus signs respectively refer to positrons and electrons). The evolution of the proper number density $n_\pm$ is governed by the continuity equation, which is
\begin{equation}
\label{eq:cont}
\frac{\pa}{\pa t}\left(\gamma_\pm n_\pm\right) + \nabla\cdot\left(n_\pm {\bm u}_\pm\right)=0 \;.
\end{equation}
The evolution of the wave vector potential is governed by Amp\`ere's law, which is
\begin{equation}
\label{eq:ampere}
\frac{\pa^2\tilde{\bm a}}{\pa t^2} -\nabla^2\tilde{\bm a} = \frac{4\pi e^2}{m}\left(n_+{\bm u}_+ - n_-{\bm u}_- \right) \;.
\end{equation}
In Sects.~\ref{sec:regime}-\ref{sec:set}, we manipulate Eqs.~\eqref{eq:euler}-\eqref{eq:ampere} to derive the equations that govern the evolution of the wave envelope.

\subsection{Weakly non-linear regime}
\label{sec:regime}

When the Lorentz factor of the particles is nearly constant inside the wave packet (i.e.,~$\gamma_\pm\simeq\gamma_0$ in the wave frame), the non-linear terms of Eqs.~\eqref{eq:euler}-\eqref{eq:ampere} can be treated as a small perturbation, because the current $n_\pm{\bm u}_\pm$ is roughly proportional to $\tilde{\bm a}$ \citep{Sobacchi+2024b, Sobacchi+2024a}. The key condition $\gamma_\pm\simeq\gamma_0$ is satisfied when $a_0\ll \omega_0/\max[\omega_{\rm P}, \omega_{\rm L}]$ \citep{Sobacchi+2024b}. In the weakly non-linear regime, high-order harmonics of the fields can be neglected. Second-order harmonics would be important to calculate non-linear corrections to the dispersion relation. However, these corrections do not affect the filamentation of electromagnetic waves \citep{Sobacchi+2024a}. Then, the vector potential can be presented as
\begin{equation}
\label{eq:afast}
\tilde{\bm a} = {\bm \psi}_0 + \frac{1}{2}a {\bm e}_y{\rm e}^{-{\rm i}\omega t}+\ldots+{\rm c.c.}\;,
\end{equation}
where c.c.~indicates the complex conjugate of the oscillating term proportional to ${\rm e}^{-{\rm i}\omega t}$, and the dots indicate high-order harmonics. The complex function $a ({\bm x}, t)$ describes the envelope of the wave packet. The strength parameter of the wave is $a_0=\max[|a|]$. We emphasize that we do not assume $a_0\ll 1$. The real function ${\bm \psi}_0 ({\bm x}, t)$ describes the secular evolution of the vector potential. We will show that ${\bm \psi}_0$ does not vanish because the electromagnetic wave compresses the plasma inside the wave packet and consequently amplifies the background magnetic field. The functions $a$ and ${\bm \psi}_0$ vary on spatial and temporal scales $\gg 1/\omega$.

As discussed earlier, the particles ahead of the wave packet have a large four-velocity, $u_0=k_0/\omega$, in the wave frame. Since we are interested in the weakly non-linear regime where the Lorentz factor of the particles inside the wave packet is nearly constant, the four-velocity can be presented as
\begin{equation}
\label{eq:ufast}
{\bm u}_\pm = -u_0{\bm e}_z + \delta{\bm u}_{0\pm} + \frac{1}{2}{\bm u}_{\rm f\pm}{\rm e}^{-{\rm i}\omega t}+\ldots+{\rm c.c.}\;,
\end{equation}
where the real function $\delta {\bm u}_{0\pm}$ describes the secular evolution of the four-velocity, and the complex function ${\bm u}_{\rm f\pm}$ describes the envelope of the fast oscillations.

In the following, we study separately the fast oscillations of the physical quantities, which occur on the time scale $1/\omega$, and their secular evolution. We retain secular terms of the order of $|\delta \bm{u}_{0\pm}|/\gamma_0$ and $|\bm{u}_{\rm f\pm}|^2/\gamma_0^2$, and oscillating terms of the order of $|{\bm u}_{{\rm f}\pm}||\delta \bm{u}_{0\pm}|/\gamma_0^2$. We neglect oscillating terms of the order of $|\bm{u}_{\rm f\pm}|^3/\gamma_0^3$, which do not affect the filamentation of electromagnetic waves \citep[][]{Sobacchi+2024a}. The Lorentz factor can be approximated as
\begin{equation}
\label{eq:gammafast}
\frac{\gamma_\pm}{\gamma_0} = 1 -\frac{u_0}{\gamma_0}\frac{\delta u_{0z\pm}}{\gamma_0} + \frac{\left|{\bm u}_{\rm f\pm}\right|^2}{4\gamma_0^2} -\frac{u_0^2}{\gamma_0^2}\frac{\left| u_{{\rm f}z\pm}\right|^2}{4\gamma_0^2}
-\left[\frac{u_0}{\gamma_0}\frac{u_{{\rm f}z\pm}}{2\gamma_0} \left(1+ \frac{u_0}{\gamma_0}\frac{\delta u_{0z\pm}}{\gamma_0} \right) - \frac{{\bm u}_{{\rm f}\pm}\cdot\delta{\bm u}_{0\pm}}{2\gamma_0^2} \right] {\rm e}^{-{\rm i}\omega t} +\ldots+{\rm c.c.}
\end{equation}
From Eqs.~\eqref{eq:ufast}-\eqref{eq:gammafast}, one finds
\begin{align}
\nonumber
\frac{{\bm u}_\pm}{\gamma_\pm} = & -\frac{u_0}{\gamma_0} \left( 1+ \frac{u_0}{\gamma_0}\frac{\delta u_{0z\pm}}{\gamma_0} - \frac{\left|{\bm u}_{\rm f\pm}\right|^2}{4\gamma_0^2} + \frac{u_0^2}{\gamma_0^2}\frac{3\left|u_{{\rm f}z\pm}\right|^2}{4\gamma_0^2} \right) {\bm e}_z + \frac{\delta {\bm u}_{0\pm}}{\gamma_0} +\frac{u_0}{\gamma_0}\frac{u_{{\rm f}z\pm} {\bm u}^*_{\rm f\pm} + u_{{\rm f}z\pm}^* {\bm u}_{\rm f\pm}}{4\gamma_0^2}
+ \\
\label{eq:vfast}
& + \left[ \frac{{\bm u}_{\rm f\pm}}{2\gamma_0}\left(1 +\frac{u_0}{\gamma_0}\frac{\delta u_{0z\pm}}{\gamma_0} \right)
+\frac{u_0}{\gamma_0}\frac{u_{{\rm f}z\pm}}{2\gamma_0}\frac{\delta {\bm u}_{0\pm}}{\gamma_0} \right] {\rm e}^{-{\rm i}\omega t} - \left[ \frac{u_0^2}{\gamma_0^2} \frac{u_{{\rm f}z\pm}}{2\gamma_0} \left( 1 + \frac{u_0}{\gamma_0}\frac{3\delta u_{0z\pm}}{\gamma_0} \right) -\frac{u_0}{\gamma_0} \frac{{\bm u}_{{\rm f}\pm}\cdot\delta{\bm u}_{0\pm}}{2\gamma_0^2}
 \right] {\bm e}_z \; {\rm e}^{-{\rm i}\omega t} + \ldots + {\rm c.c.} \;,
\end{align}
where ${\bm u}^*_{\rm f\pm}$ denotes the complex conjugate of ${\bm u}_{\rm f\pm}$.

\subsection{Fast oscillations}
\label{sec:fast}

In the following, we study the fast oscillations of the physical quantities on the time scale $1/\omega$. We exploit the fact that the spatial derivatives do not affect the fast oscillations because the wave vector vanishes in the wave frame. The proper number density can be presented as
\begin{equation}
\frac{n_\pm}{n_0} =1+\frac{\delta n_{0\pm}}{n_0} + \frac{n_{{\rm f}\pm}}{2n_0} {\rm e}^{-{\rm i}\omega t}+\ldots+{\rm\; c.c.}\;,
\end{equation}
where the real function $\delta n_{0\pm}$ describes the secular evolution of the proper number density, and the complex function $n_{{\rm f}\pm}$ describes the envelope of the fast oscillations. The function $n_{{\rm f}\pm}$ can be determined from Eq.~\eqref{eq:cont}. Since spatial derivatives do not affect fast oscillations, the oscillating terms of $\gamma_\pm n_\pm$ should vanish. Using Eq.~\eqref{eq:gammafast} to eliminate $\gamma_\pm$, one finds
\begin{equation}
\label{eq:nfast}
\frac{n_\pm}{n_0} = 1+\frac{\delta n_{0\pm}}{n_0} + \left[ \frac{u_0}{\gamma_0}\frac{u_{{\rm f}z\pm}}{2\gamma_0} \left(1 +\frac{\delta n_{0\pm}}{n_0} + \frac{u_0}{\gamma_0}\frac{2\delta u_{0z\pm}}{\gamma_0} \right) - \frac{{\bm u}_{{\rm f}\pm}\cdot\delta{\bm u}_{0\pm}}{2\gamma_0^2} \right] {\rm e}^{-{\rm i}\omega t} +\ldots+{\rm c.c.}
\end{equation}
From Eqs.~\eqref{eq:gammafast} and \eqref{eq:nfast}, one finds
\begin{equation}
\label{eq:rho}
\frac{\gamma_\pm n_\pm}{\gamma_0n_0} = 1 +\frac{\delta n_{0\pm}}{n_0} - \frac{u_0}{\gamma_0}\frac{\delta u_{0z\pm}}{\gamma_0} + \frac{\left|{\bm u}_{{\rm f}\pm}\right|^2}{4\gamma_0^2} -\frac{u_0^2}{\gamma_0^2}\frac{3\left|u_{{\rm f}z\pm}\right|^2}{4\gamma_0^2} +\ldots +{\rm c.c.}
\end{equation}
From Eqs.~\eqref{eq:ufast} and \eqref{eq:nfast}, one finds
\begin{align}
\nonumber
\frac{n_\pm{\bm u}_\pm}{\gamma_0n_0} = -\frac{u_0}{\gamma_0} \left(1+ \frac{\delta n_{0\pm}}{n_0}\right) {\bm e}_z +\frac{\delta {\bm u}_{0\pm}}{\gamma_0} & + \frac{u_0}{\gamma_0}\frac{u_{{\rm f}z\pm} {\bm u}^*_{\rm f\pm} + u_{{\rm f}z\pm}^* {\bm u}_{\rm f\pm}}{4\gamma_0^2} + \left[ \left(1+\frac{\delta n_{0\pm}}{n_0}\right)\frac{{\bm u}_{\rm f\pm}}{2\gamma_0} + \frac{u_0}{\gamma_0}\frac{u_{{\rm f}z\pm}}{2\gamma_0} \frac{\delta {\bm u}_{0\pm}}{\gamma_0} \right]{\rm e}^{-{\rm i}\omega t}  +\\
\label{eq:j}
& - \frac{u_0}{\gamma_0} \left[ \frac{u_0}{\gamma_0}\frac{u_{{\rm f}z\pm}}{2\gamma_0} \left(1+\frac{\delta n_{0\pm}}{n_0} +\frac{u_0}{\gamma_0}\frac{2\delta u_{0z\pm}}{\gamma_0}  \right) - \frac{{\bm u}_{{\rm f}\pm}\cdot\delta{\bm u}_{0\pm}}{2\gamma_0^2}  \right] {\bm e}_z \; {\rm e}^{-{\rm i}\omega t}  +\ldots +{\rm c.c.}
\end{align}
In Eqs.~\eqref{eq:nfast} and \eqref{eq:j}, we retained oscillating terms of the order of $|{\bm u}_{{\rm f}\pm}||\delta \bm{u}_{0\pm}|/\gamma_0^2$ and $|{\bm u}_{{\rm f}\pm}||\delta n_{0\pm}|/\gamma_0 n_0$.

\subsubsection{Dispersion relation}
\label{sec:DR}

In the following, we determine the dispersion relation of the wave at leading order (i.e.,~we neglect the non-linear terms). These terms are important for the evolution of the wave envelope, which is studied in Sect.~\ref{sec:envelope}. First, one should express the fast oscillations of the four-velocity as a function of $a$. Substituting Eqs.~\eqref{eq:afast}, \eqref{eq:ufast}, \eqref{eq:vfast} into Eq.~\eqref{eq:euler}, neglecting the spatial derivatives, one finds
\begin{equation}
\label{eq:ufastyz}
\left(1-\frac{\omega_{\rm L}^2}{\gamma_0^2\omega^2}\right) u_{{\rm f}y\pm} \simeq \mp a  \;,\qquad \left(1-\frac{\omega_{\rm L}^2}{\gamma_0^2\omega^2}\right) u_{{\rm f}z\pm} \simeq {\rm i}\frac{\omega_{\rm L}}{\omega} a \;.
\end{equation}
Then, one should substitute Eqs.~\eqref{eq:afast} and \eqref{eq:j} into Eq.~\eqref{eq:ampere}. On the left-hand side and on the right-hand side, neglecting the non-linear terms, one can approximate respectively $\pa^2\tilde{\bm a}/\pa t^2-\nabla^2\tilde{\bm a} \simeq -\omega^2a{\bm e}_y$ and $n_+{\bm u}_+-n_-{\bm u}_-\simeq 2n_0u_{{\rm f}y+}{\bm e}_y$, where $u_{{\rm f}y+}$ is given by Eq.~\eqref{eq:ufastyz}. Then, the dispersion relation is given by
\begin{equation}
\label{eq:DRwave}
\left(1-\frac{\omega_{\rm L}^2}{\gamma_0^2\omega^2}\right) \omega^2 = \omega_{\rm P}^2 \;.
\end{equation}
Taking into account that $\omega^2=\omega_0^2-k_0^2$ and $\gamma_0^2\omega^2=\omega_0^2$, one sees that Eq.~\eqref{eq:DRwave} is equivalent to the standard dispersion relation in the wind frame, which is $(\omega_0^2-k_0^2) (1-\omega_{\rm L}^2/\omega_0^2)=\omega_{\rm P}^2$. For $\omega_0\gg\omega_{\rm L}$, Eq.~\eqref{eq:DRwave} implies $\omega\simeq\omega_{\rm P}$, and therefore $\gamma_0\simeq\omega_0/\omega_{\rm P}$.

To derive the dispersion relation, we assumed that the Lorentz factor of the particles inside the wave packet is nearly constant (i.e.,~$\gamma_\pm\simeq\gamma_0$). Eq.~\eqref{eq:gammafast} shows that the Lorentz factor is nearly constant for $u_{{\rm f}y\pm}\ll\gamma_0$ and $u_{{\rm f}z\pm}\ll\gamma_0$. Since for $\omega_0\gg\omega_{\rm L}$ Eq.~\eqref{eq:ufastyz} gives $u_{{\rm f}\pm y}\simeq \mp a$ and $u_{{\rm f}\pm z}\simeq {\rm i} (\omega_{\rm L}/\omega_{\rm P}) a$, the condition $\gamma_\pm \simeq\gamma_0$ is satisfied for
\begin{equation}
\label{eq:cond}
a_0\ll \frac{\omega_0}{\max\left[\omega_{\rm P},\omega_{\rm L}\right]} \;,
\end{equation}
where $a_0=\max[|a|]$. Eq.~\eqref{eq:cond} has a simple physical interpretation \citep{Sobacchi+2024b}. The condition $a_0\ll \omega_0/\omega_{\rm P}$ implies that in the wind frame the longitudinal velocity of the particles inside the wave packet should be smaller than the group velocity. The condition $a_0\ll \omega_0/\omega_{\rm L}$ implies that the background magnetic field inside the wave packet, which is amplified by a factor of order $a_0^2$ in the wind frame, should be smaller than the wave field. We emphasize that Eq.~\eqref{eq:cond} is always satisfied by electromagnetic precursors in relativistic shocks. Taking into account that $\omega_0$ and $a_0$ are respectively given by Eqs.~\eqref{eq:omegaprec} and \eqref{eq:aprec}, one finds $a_0 \max[\omega_{\rm P},\omega_{\rm L}]/\omega_0=a_0\omega_{\rm P}\max[1,\sqrt{\sigma}]/\omega_0\simeq 0.01$.

\subsubsection{Evolution of the wave envelope}
\label{sec:envelope}

The equation that governs the evolution of the wave envelope on time scales $\gg 1/\omega$ can be determined by substituting Eqs.~\eqref{eq:afast} and \eqref{eq:j} into the $y$ component of Eq.~\eqref{eq:ampere}. Considering the resonant terms (i.e.,~the terms proportional to ${\rm e}^{-{\rm i}\omega t}$, or ${\rm e}^{{\rm i}\omega t}$), one finds
\begin{equation}
\label{eq:waveenvaux}
\frac{\pa^2 a}{\pa t^2}-2{\rm i}\omega\frac{\pa a}{\pa t} -\omega^2 a -\nabla^2 a  = \frac{\omega_{\rm P}^2}{2} \left[ \frac{\delta n_{0+}}{n_0} u_{{\rm f}y+} - \frac{\delta n_{0-}}{n_0} u_{{\rm f}y-} +\frac{u_0}{\gamma_0} \left(\frac{\delta u_{0y+}}{\gamma_0}u_{{\rm f}z+} - \frac{\delta u_{0y-}}{\gamma_0} u_{{\rm f}z-} \right) \right] + \frac{\omega_{\rm P}^2}{2} \left( u_{{\rm f}y+}-u_{{\rm f}y-} \right) \;.
\end{equation}
The first term on the left-hand side of Eq.~\eqref{eq:waveenvaux} can be neglected because $\pa^2 a/\pa t^2\ll\omega \pa a/\pa t$. In the first term on the right-hand side, one can approximate $u_{{\rm f}y\pm}\simeq\mp a$ and $u_{{\rm f}z\pm}\simeq {\rm i}(\omega_{\rm L}/\omega_{\rm P})a$, which follow from Eq.~\eqref{eq:ufastyz} since $\max [\omega_{\rm P},\omega_{\rm L}]\ll\omega_0$. In the second term, one should retain the non-linear corrections to $u_{{\rm f}y+}-u_{{\rm f}y-}$. Substituting Eqs.~\eqref{eq:afast}, \eqref{eq:ufast}, \eqref{eq:vfast} into Eq.~\eqref{eq:euler}, one finds
\begin{equation}
\label{eq:ufastnl}
\left(1-\frac{\omega_{\rm L}^2}{\gamma_0^2\omega^2}\right) \left( u_{{\rm f}y+}-u_{{\rm f}y-} \right) = -2 a - {\rm i} \frac{\omega_{\rm L}}{\omega} \left( \frac{\delta u_{0y+}}{\gamma_0} - \frac{\delta u_{0y-}}{\gamma_0}  \right) a \;.
\end{equation}
On the right-hand side of Eq.~\eqref{eq:ufastnl}, we approximated $u_0\simeq\gamma_0$, as appropriate because $\gamma_0\gg 1$. Substituting Eq.~\eqref{eq:ufastnl} into Eq.~\eqref{eq:waveenvaux}, and using Eq.~\eqref{eq:DRwave} to eliminate $\omega$, one finds
\begin{equation}
\label{eq:envaux}
\frac{\rm i}{\omega_{\rm P}}\frac{\pa a}{\pa t} = -\frac{1}{2\omega_{\rm P}^2}\nabla^2a +\frac{1}{2}\delta\tilde{\rho} a \;,
\end{equation}
where we defined
\begin{equation}
\delta\tilde{\rho} = \frac{1}{2}\left(\frac{\delta n_{0+}}{n_0}+\frac{\delta n_{0-}}{n_0}\right) \;.
\end{equation}
The evolution of the wave envelope is governed by Eq.~\eqref{eq:envaux}.

\subsection{Secular evolution}
\label{sec:secular}

Studying the secular evolution of the physical quantities is key for the wave envelope, as Eq.~\eqref{eq:envaux} depends on the secular variable $\delta \tilde{\rho}$. One should substitute Eqs.~\eqref{eq:afast}-\eqref{eq:j} into Eqs.~\eqref{eq:euler}-\eqref{eq:ampere}, and average over the fast oscillations. We start by considering Amp\`ere's law. Substituting Eqs.~\eqref{eq:afast} and \eqref{eq:j} into Eq.~\eqref{eq:ampere}, and taking into account that $u_{{\rm f}y\pm}\simeq \mp a$ and $u_{{\rm f}z\pm} \simeq {\rm i} (\omega_{\rm L}/\omega_{\rm P})a$, which follow from Eq.~\eqref{eq:ufastyz}, one finds
\begin{equation}
\label{eq:psi0}
\frac{\pa^2}{\pa t^2}\frac{{\bm\psi}_0}{\gamma_0} - \nabla^2\frac{{\bm\psi}_0}{\gamma_0} = \frac{\omega_{\rm P}^2}{2} \left[\frac{\delta{\bm u}_{0+}}{\gamma_0} -\frac{\delta{\bm u}_{0-}}{\gamma_0} -\frac{u_0}{\gamma_0}\left(\frac{\delta n_{0+}}{n_0} - \frac{\delta n_{0-}}{n_0} \right){\bm e}_z \right] \;.
\end{equation}
The transverse component of Eq.~\eqref{eq:psi0} can be presented as
\begin{equation}
\label{eq:deltauperp}
\frac{\delta{\bm u}_{0\perp +}}{\gamma_0} - \frac{\delta{\bm u}_{0\perp -}}{\gamma_0} = \frac{2}{\omega_{\rm P}^2} \left( \frac{\pa^2}{\pa t^2}\frac{{\bm\psi}_{0\perp}}{\gamma_0}  -\nabla^2\frac{{\bm\psi}_{0\perp}}{\gamma_0}\right) \;,
\end{equation}
where we defined $\delta{\bm u}_{0\perp\pm}=\delta u_{0\pm x}{\bm e}_x+\delta u_{0\pm y}{\bm e}_y$, and ${\bm \psi}_{0\perp}=\psi_{0x}{\bm e}_x+\psi_{0y}{\bm e}_y$. Now we consider the particles equation of motion. Substituting Eqs.~\eqref{eq:afast}-\eqref{eq:vfast} into Eq.~\eqref{eq:euler}, and taking into account that $u_{{\rm f}y\pm}\simeq \mp a$ and $u_{{\rm f}z\pm} \simeq {\rm i} (\omega_{\rm L}/\omega_{\rm P})a$, one finds
\begin{align}
\nonumber
\frac{\pa}{\pa t} \left(\frac{\delta{\bm u}_{0\pm}}{\gamma_0}\pm\frac{{\bm \psi}_0}{\gamma_0} \right) + \frac{u_0}{\gamma_0} {\bm e}_z\times\left[\nabla\times \left(\frac{\delta{\bm u}_{0\pm}}{\gamma_0}\pm\frac{{\bm \psi}_0}{\gamma_0}\right) \right] & \pm {\rm i}\frac{\omega_{\rm L}}{\omega_{\rm P}}\left[ \frac{a}{2\gamma_0}\frac{\pa}{\pa y}\left(\frac{a^*}{2\gamma_0}\right) - \frac{a^*}{2\gamma_0}\frac{\pa}{\pa y}\left(\frac{a}{2\gamma_0}\right) \right] {\bm e}_z = \\
\label{eq:u0}
& = -\nabla\left( \frac{\left|a\right|^2}{4\gamma_0^2} -\frac{u_0}{\gamma_0}\frac{\delta u_{0z\pm}}{\gamma_0} \right)  \pm \left( \frac{\delta{\bm u}_{0\pm}}{\gamma_0} -\frac{u_0^2}{\gamma_0^2}\frac{\delta u_{0z\pm}}{\gamma_0}{\bm e}_z + \frac{\left|a\right|^2}{4\gamma_0^2}{\bm e}_z \right) \times \omega_{\rm L}{\bm e}_x \;,
\end{align}
where $a^*$ denotes the complex conjugate of $a$. In the derivation of Eq.~\eqref{eq:u0}, we neglected terms of order $a_0^2/\gamma_0^4$. It is convenient to present two separate equations for the variables $\delta{\bm u}_{0+}+\delta{\bm u}_{0-}$ and $\delta{\bm u}_{0+}-\delta{\bm u}_{0-}$. As we demonstrate in Appendix \ref{sec:motion}, the equation for the evolution of $\delta{\bm u}_{0+}+\delta{\bm u}_{0-}$ is
\begin{equation}
\label{eq:v0}
\frac{\pa}{\pa t} \left( \frac{\delta {\bm u}_{0+}}{\gamma_0} + \frac{\delta {\bm u}_{0-}}{\gamma_0} \right) -\frac{u_0}{\gamma_0}\frac{\pa}{\pa z} \left( \frac{\delta {\bm u}_{0+}}{\gamma_0} + \frac{\delta {\bm u}_{0-}}{\gamma_0} \right) = -\nabla\frac{\left|a\right|^2}{2\gamma_0^2}  + \frac{1}{\gamma_0^2}\left(\frac{\delta u_{0z+}}{\gamma_0}-\frac{\delta u_{0z+}}{\gamma_0}\right)\omega_{\rm L}{\bm e}_y + \frac{2\omega_{\rm L}}{\omega_{\rm P}^2} \left( \nabla^2\frac{{\psi}_{0y}}{\gamma_0} -\frac{\pa^2}{\pa t^2} \frac{{\psi}_{0y}}{\gamma_0} \right) {\bm e}_z \;.
\end{equation}
The first term on the right-hand side of Eq.~\eqref{eq:v0} is the ponderomotive force. The equation for the evolution of $\delta{\bm u}_{0+}-\delta{\bm u}_{0-}$ is
\begin{align}
\nonumber
\frac{\pa}{\pa t} \left(\frac{\delta {\bm u}_{0+}}{\gamma_0} - \frac{\delta {\bm u}_{0-}}{\gamma_0} +\frac{2{\bm\psi}_0}{\gamma_0} \right) & -\frac{u_0}{\gamma_0}\frac{\pa}{\pa z} \left( \frac{\delta {\bm u}_{0+}}{\gamma_0} - \frac{\delta {\bm u}_{0-}}{\gamma_0} +\frac{2{\bm\psi}_0}{\gamma_0} \right) +\frac{u_0}{\gamma_0}\nabla\frac{2\psi_{0z}}{\gamma_0} + {\rm i}\frac{2\omega_{\rm L}}{\omega_{\rm P}}\left[ \frac{a}{2\gamma_0}\frac{\pa}{\pa y}\left(\frac{a^*}{2\gamma_0}\right) - \frac{a^*}{2\gamma_0}\frac{\pa}{\pa y}\left(\frac{a}{2\gamma_0}\right) \right] {\bm e}_z = \\
\label{eq:B0}
& = -\left(\frac{\delta u_{0y+}}{\gamma_0}+\frac{\delta u_{0y-}}{\gamma_0}\right)\omega_{\rm L}{\bm e}_z  + \frac{1}{\gamma_0^2}\left(\frac{\delta u_{0z+}}{\gamma_0}+\frac{\delta u_{0z-}}{\gamma_0}\right)\omega_{\rm L}{\bm e}_y + \frac{\left|a\right|^2}{2\gamma_0^2}\omega_{\rm L}{\bm e}_y \;.
\end{align}
Finally, we consider the continuity equation. Substituting Eqs.~\eqref{eq:rho}-\eqref{eq:j} into Eq.~\eqref{eq:cont}, and taking into account that $u_{{\rm f}y\pm}\simeq \mp a$ and $u_{{\rm f}z\pm} \simeq {\rm i} (\omega_{\rm L}/\omega_{\rm P})a$, one finds
\begin{equation}
\label{eq:cont0aux}
\frac{\pa}{\pa t}\left(\frac{\delta n_{0\pm}}{n_0} -\frac{u_0}{\gamma_0}\frac{\delta u_{0z\pm}}{\gamma_0} +\frac{\left|a\right|^2}{4\gamma_0^2} - \frac{\omega_{\rm L}^2}{\omega_{\rm P}^2}\frac{\left|a\right|^2}{2\gamma_0^2} \right) + \nabla \cdot \left( \frac{\delta{\bm u}_{0\pm}}{\gamma_0} -\frac{u_0}{\gamma_0}\frac{\delta n_{0\pm}}{n_0}{\bm e}_z +\frac{\omega_{\rm L}^2}{\omega_{\rm P}^2}\frac{\left|a\right|^2}{2\gamma_0^2}{\bm e}_z \right) = 0 \;.
\end{equation}
In the derivation of Eq.~\eqref{eq:cont0aux}, we neglected terms of order $a_0^2/\gamma_0^4$. It is convenient to present an equation for the variable $\delta \tilde{\rho}$, which appears in Eq.~\eqref{eq:envaux}. One finds
\begin{equation}
\label{eq:cont0}
\frac{\pa}{\pa t} \left(2\delta\tilde{\rho} -\frac{u_0}{\gamma_0}\frac{\delta u_{0z+}}{\gamma_0} -  \frac{u_0}{\gamma_0}\frac{\delta u_{0z-}}{\gamma_0} + \frac{\left|a\right|^2}{2\gamma_0^2} - \frac{\omega_{\rm L}^2}{\omega_{\rm P}^2}\frac{\left|a\right|^2}{\gamma_0^2} \right)
-2\frac{u_0}{\gamma_0}\frac{\pa\delta\tilde{\rho}}{\pa z} + \frac{\pa}{\pa z}\left(\frac{\omega_{\rm L}^2}{\omega_{\rm P}^2}\frac{\left|a\right|^2}{\gamma_0^2}\right) +\nabla\cdot \left(\frac{\delta{\bm u}_{0+}}{\gamma_0}+\frac{\delta{\bm u}_{0-}}{\gamma_0}\right) = 0 \;.
\end{equation}
In the following, we make some algebraic manipulations to simplify our set of equations. In Sect.~\ref{sec:B}, we study the evolution of the background electromagnetic fields. Then, in Sect.~\ref{sec:fluid}, we study the velocity and density of the fluid.

\subsubsection{Background electromagnetic fields}
\label{sec:B}

The background electromagnetic fields are described by the function ${\bm \psi}_0$. To study the secular evolution of the magnetic field, it is convenient to introduce the variable $\delta\bm{\omega}_{\rm L}=\nabla\times\bm{\psi}_0$, or equivalently
\begin{equation}
\label{eq:defomegaL}
\delta \omega_{{\rm L}x} = \frac{\pa\psi_{0z}}{\pa y} - \frac{\pa\psi_{0y}}{\pa z}\;, \qquad \delta \omega_{{\rm L}y} = \frac{\pa\psi_{0x}}{\pa z} - \frac{\pa\psi_{0z}}{\pa x} \;, \qquad \delta \omega_{{\rm L}z} = \frac{\pa\psi_{0y}}{\pa x} - \frac{\pa\psi_{0x}}{\pa y} \;.
\end{equation}
Since $\nabla\cdot\delta\bm{\omega}_{\rm L}=0$, one has
\begin{equation}
\label{eq:divB}
\frac{\pa\delta\omega_{{\rm L}z}}{\pa z} = -\frac{\pa\delta\omega_{{\rm L}x}}{\pa x} -\frac{\pa\delta\omega_{{\rm L}y}}{\pa y} \;.
\end{equation}
To study the secular evolution of the electric field, it is convenient to introduce the variable $\delta\bm{\omega}_{\rm E}=-\pa\bm{\psi}_0/\pa t$, or equivalently
\begin{equation}
\label{eq:defomegaE}
\delta \omega_{{\rm E}x} = - \frac{\pa\psi_{0x}}{\pa t}\;, \qquad \delta \omega_{{\rm E}y} = - \frac{\pa\psi_{0y}}{\pa t} \;, \qquad \delta \omega_{{\rm E}z} = - \frac{\pa\psi_{0z}}{\pa t} \;.
\end{equation}
Since $\nabla\cdot\bm{\psi}_0=0$ in the Coulomb gauge, one has
\begin{equation}
\label{eq:divE}
\frac{\pa\delta\omega_{{\rm E}z}}{\pa z} = -\frac{\pa\delta\omega_{{\rm E}x}}{\pa x} -\frac{\pa\delta\omega_{{\rm E}y}}{\pa y} \;.
\end{equation}
From Eqs.~\eqref{eq:defomegaL} and \eqref{eq:defomegaE}, one finds
\begin{equation}
\label{eq:curlE}
\frac{\pa\delta\omega_{{\rm L}x}}{\pa t} = \frac{\pa\delta\omega_{{\rm E}y}}{\pa z}-\frac{\pa\delta\omega_{{\rm E}z}}{\pa y} \;,\qquad \frac{\pa\delta\omega_{{\rm L}y}}{\pa t} = \frac{\pa\delta\omega_{{\rm E}z}}{\pa x}-\frac{\pa\delta\omega_{{\rm E}x}}{\pa z}\;.
\end{equation}
It is also convenient to introduce the variable
\begin{equation}
\delta {\bf v} = \frac{1}{2} \left( \frac{\delta {\bm u}_{0+}}{\gamma_0} + \frac{\delta {\bm u}_{0-}}{\gamma_0} \right) \;.
\end{equation}

First, we consider the $x$ component of Eq.~\eqref{eq:B0}. Using Eq.~\eqref{eq:deltauperp} to eliminate $\delta u_{0+x}-\delta u_{0-x}$, and taking into account that $\psi_{0x}$ varies on spatial and temporal scales $\gg 1/\omega_{\rm P}$, one sees that terms proportional to $\delta u_{0+x}-\delta u_{0-x}$ can be neglected. Then, the $x$ component of Eq.~\eqref{eq:B0} can be approximated as
\begin{equation}
\label{eq:Ex}
\delta\omega_{{\rm E}x}=-\frac{u_0}{\gamma_0}\delta\omega_{{\rm L}y}\;.
\end{equation}
Substituting Eq.~\eqref{eq:Ex} into Eq.~\eqref{eq:curlE}, one finds
\begin{equation}
\label{eq:Byaux}
\frac{\pa\delta\omega_{{\rm L}y}}{\pa t} - \frac{u_0}{\gamma_0} \frac{\pa\delta\omega_{{\rm L}y}}{\pa z} = \frac{\pa\delta\omega_{{\rm E}z}}{\pa x} \;.
\end{equation}
Next, we consider the $y$ component of Eq.~\eqref{eq:B0}. As we discussed in the derivation of Eq.~\eqref{eq:Ex}, terms proportional to $\delta u_{0+y}-\delta u_{0-y}$ can be neglected. Then, one can approximate
\begin{equation}
\label{eq:Ey}
\delta\omega_{{\rm E}y} =  \frac{u_0}{\gamma_0} \delta\omega_{{\rm L}x} - \frac{\left|a\right|^2}{4\gamma_0}\omega_{\rm L}-\frac{\delta v_z}{\gamma_0}\omega_{\rm L}\;.
\end{equation}
Substituting Eq.~\eqref{eq:Ey} into Eq.~\eqref{eq:curlE}, one finds
\begin{equation}
\label{eq:Bxaux}
\frac{\pa\delta\omega_{{\rm L}x}}{\pa t} - \frac{\pa}{\pa z} \left( \frac{u_0}{\gamma_0} \delta\omega_{{\rm L}x} - \frac{\left|a\right|^2}{4\gamma_0}\omega_{\rm L} - \frac{\delta v_z}{\gamma_0} \omega_{\rm L} \right) = -\frac{\pa\delta\omega_{{\rm E}z}}{\pa y} \;.
\end{equation}
Substituting Eqs.~\eqref{eq:Ex} and \eqref{eq:Ey} into Eq.~\eqref{eq:divE}, one finds
\begin{equation}
\label{eq:divEaux}
\frac{\pa\delta\omega_{{\rm E}z}}{\pa z} = \frac{u_0}{\gamma_0} \frac{\pa\delta\omega_{{\rm L}y}}{\pa x} - \frac{\pa}{\pa y}\left(\frac{u_0}{\gamma_0} \delta\omega_{{\rm L}x} - \frac{\left|a\right|^2}{4\gamma_0}\omega_{\rm L} - \frac{\delta v_z}{\gamma_0} \omega_{\rm L}\right) \;.
\end{equation}
Eqs.~\eqref{eq:divB} and \eqref{eq:divEaux} can be presented in a more convenient form. Applying the operator $\pa/\pa t-(u_0/\gamma_0)\pa/\pa z$ to both sides of Eq.~\eqref{eq:divB}, and using Eqs.~\eqref{eq:Byaux} and \eqref{eq:Bxaux} to eliminate $\delta\omega_{{\rm L}x}$ and $\delta\omega_{{\rm L}y}$, one finds
\begin{equation}
\frac{\pa}{\pa z}\left( \frac{\pa\delta\omega_{{\rm L}z}}{\pa t} - \frac{u_0}{\gamma_0} \frac{\pa\delta\omega_{{\rm L}z}}{\pa z} \right) = \frac{\pa^2}{\pa x\pa z}\left( \frac{\left|a\right|^2}{4\gamma_0}\omega_{\rm L}+ \frac{\delta v_z}{\gamma_0}\omega_{\rm L} \right) \;,
\end{equation}
which implies
\begin{equation}
\label{eq:divBfinal}
\frac{\pa\delta\omega_{{\rm L}z}}{\pa t} - \frac{u_0}{\gamma_0} \frac{\pa\delta\omega_{{\rm L}z}}{\pa z}  = \frac{\pa}{\pa x}\left( \frac{\left|a\right|^2}{4\gamma_0}\omega_{\rm L}+ \frac{\delta v_z}{\gamma_0}\omega_{\rm L} \right) \;.
\end{equation}
Similarly, applying the operator $\pa/\pa t-(u_0/\gamma_0) \pa/\pa z$ to both sides of Eq.~\eqref{eq:divEaux}, one finds
\begin{equation}
\label{eq:divEfinal}
\frac{u_0}{\gamma_0} \left(\frac{\pa^2\delta\omega_{{\rm E}z}}{\pa x^2} + \frac{\pa^2\delta\omega_{{\rm E}z}}{\pa y^2} + \frac{\pa^2\delta\omega_{{\rm E}z}}{\pa z^2} \right) = \frac{\pa}{\pa t} \left[\frac{\pa\delta\omega_{{\rm E}z}}{\pa z} -\frac{\pa}{\pa y} \left( \frac{\left|a\right|^2}{4\gamma_0}\omega_{\rm L}+ \frac{\delta v_z}{\gamma_0}\omega_{\rm L} \right) \right] \;.
\end{equation}
The evolution of $\delta\omega_{{\rm L}x}$, $\delta\omega_{{\rm L}y}$, $\delta\omega_{{\rm L}z}$, $\delta\omega_{{\rm E}z}$ is governed by Eqs.~\eqref{eq:Byaux}, \eqref{eq:Bxaux}, \eqref{eq:divBfinal}, \eqref{eq:divEfinal}. After determining the solution of these equations, one can calculate $\delta\omega_{{\rm E}x}$ from Eq.~\eqref{eq:Ex}, and $\delta\omega_{{\rm E}y}$ from Eq.~\eqref{eq:Ey}.

\subsubsection{Velocity and density of the fluid}
\label{sec:fluid}

The velocity and density of the fluid are described by the functions $\delta\tilde{\rho}$ and $\delta{\bm v}$. The transverse component of Eq.~\eqref{eq:v0} gives
\begin{equation}
\label{eq:deltavperp}
\frac{\pa\delta{\bm v}_\perp}{\pa t} - \frac{u_0}{\gamma_0} \frac{\pa\delta{\bm v}_\perp}{\pa z} = -\nabla_\perp\frac{\left|a\right|^2}{4\gamma_0^2} +\frac{\omega_{\rm L}}{\gamma_0^2}\delta j_z{\bm e}_y\;,
\end{equation}
where we defined
\begin{equation}
\delta j_z = \frac{1}{2} \left( \frac{\delta u_{0z+}}{\gamma_0} - \frac{\delta u_{0z-}}{\gamma_0} \right) \;.
\end{equation}
Using Eq.~\eqref{eq:identity} to simplify the right-hand side, the $z$ component of Eq.~\eqref{eq:v0} gives
\begin{equation}
\label{eq:deltavzaux} 
\frac{\pa}{\pa t} \left( \delta v_z + \frac{\omega_{\rm L}^2}{\omega_{\rm P}^2}\frac{\left|a\right|^2}{4\gamma_0^2} \right)
-\frac{u_0}{\gamma_0}\frac{\pa}{\pa z} \left( \delta v_z - \frac{\left|a\right|^2}{4\gamma_0^2} - \frac{\omega_{\rm L}^2}{\omega_{\rm P}^2}\frac{\left|a\right|^2}{4\gamma_0^2} \right) = \frac{\omega_{\rm L}^2}{\omega_{\rm P}^2}\frac{\pa}{\pa x}\left(\frac{\delta\omega_{{\rm L}z}}{\gamma_0\omega_{\rm L}}\right) - \frac{u_0}{\gamma_0} \frac{\omega_{\rm L}^2}{\omega_{\rm P}^2} \frac{\pa}{\pa y}\left(\frac{\delta\omega_{{\rm E}z}}{\gamma_0\omega_{\rm L}}\right) - \frac{\omega_{\rm L}^2}{\gamma_0^2\omega_{\rm P}^2}\frac{\pa}{\pa z}\left(\frac{\delta\omega_{{\rm L}x}}{\gamma_0\omega_{\rm L}}\right) \;.
\end{equation}
The evolution of $\delta j_z$ is governed by the $z$ component of Eq.~\eqref{eq:B0}, which gives
\begin{equation}
\label{eq:deltajz}
\frac{\pa\delta j_z}{\pa t} - \frac{u_0}{\gamma_0} \frac{\pa\delta j_z}{\pa z} = \frac{\delta\omega_{{\rm E}z}}{\gamma_0} -\omega_{\rm L} \delta v_y - {\rm i}\frac{\omega_{\rm L}}{\omega_{\rm P}}\left[ \frac{a}{2\gamma_0}\frac{\pa}{\pa y}\left(\frac{a^*}{2\gamma_0}\right) - \frac{a^*}{2\gamma_0}\frac{\pa}{\pa y}\left(\frac{a}{2\gamma_0}\right) \right] \;.
\end{equation}
The evolution of $\delta\tilde{\rho}$ is governed by Eq.~\eqref{eq:cont0}, which gives
\begin{equation}
\label{eq:deltarhoaux}
\frac{\pa}{\pa t} \left(\delta\tilde{\rho} -\frac{u_0}{\gamma_0}\delta v_z + \frac{\left|a\right|^2}{4\gamma_0^2} - \frac{\omega_{\rm L}^2}{\omega_{\rm P}^2}\frac{\left|a\right|^2}{2\gamma_0^2} \right) - \frac{\pa}{\pa z} \left( \frac{u_0}{\gamma_0} \delta\tilde{\rho} - \delta v_z - \frac{\omega_{\rm L}^2}{\omega_{\rm P}^2}\frac{\left|a\right|^2}{2\gamma_0^2}\right) = -\nabla_\perp\cdot\delta {\bm v}_\perp \;.
\end{equation}
The evolution of the fluid is governed by Eqs.~\eqref{eq:deltavperp}, \eqref{eq:deltavzaux}, \eqref{eq:deltajz}, \eqref{eq:deltarhoaux}.

\subsection{Final set of equations}
\label{sec:set}

To simplify the equations that govern the secular evolution of the system, we neglect the time derivative of $|a|$. This ``quasi-static'' approximation is extensively used to study the laser-plasma interaction \citep[e.g.,][]{Sprangle+1990}. The approximation is justified because we work in the frame that moves with the group velocity, where the time derivative of the wave envelope can be neglected. As we demonstrate in Appendix \ref{sec:finaleqs}, the equations that govern the evolution of the system can be approximated as
\begin{align}
\label{eq:final1}
\frac{\pa}{\pa t}\left(\frac{\delta \omega_{{\rm L}z}}{\gamma_0\omega_{\rm L}}\right) - \frac{\pa}{\pa z}\left(\frac{\delta \omega_{{\rm L}z}}{\gamma_0\omega_{\rm L}}\right) & = \frac{\pa}{\pa x} \frac{\left|a\right|^2}{4\gamma_0^2} \\
\label{eq:final2}
\frac{\pa\delta {\bm v}_\perp}{\pa t} - \frac{\pa\delta {\bm v}_\perp}{\pa z} & = -\nabla_\perp\frac{\left|a\right|^2}{4\gamma_0^2} +\frac{\omega_{\rm L}^2}{\gamma_0^2} \left( \frac{\delta j_z}{\omega_{\rm L}}\right) {\bm e}_y \\
\label{eq:final3}
\frac{\pa}{\pa t}\left(\frac{\delta j_z}{\omega_{\rm L}}\right) - \frac{\pa}{\pa z} \left(\frac{\delta j_z}{\omega_{\rm L}}\right) & = -\delta v_y \\
\label{eq:final4}
\frac{\pa\delta\rho}{\pa t} - \frac{\pa\delta\rho}{\pa z}  & = -\nabla_\perp\cdot\delta {\bm v}_\perp + \frac{\omega_{\rm L}^2}{\omega_{\rm P}^2}\frac{\pa}{\pa x}\left(\frac{\delta\omega_{{\rm L}z}}{\gamma_0\omega_{\rm L}}\right) \;,
\end{align}
where we defined
\begin{equation}
\label{eq:rhotildedef}
\delta\rho = \delta\tilde{\rho} - \frac{\left|a\right|^2}{4\gamma_0^2} -\frac{\omega_{\rm L}^2}{\omega_{\rm P}^2}\frac{3\left|a\right|^2}{4\gamma_0^2} \;.
\end{equation}
Since we work in the frame that moves with the group velocity of the wave packet, it is interesting to consider the case where the variables $\delta \omega_{{\rm L}z}$, $\delta {\bm v}_\perp$, $\delta j_z$, and $\delta\rho$ depend only on $z$ (and are independent of $x$, $y$, $t$). In this approximation, one would recover the results of \citet[][]{Sobacchi+2024b}.

The evolution of the wave envelope is governed by Eq.~\eqref{eq:envaux}. Since non-linear terms of order $a_0^2/\gamma_0^2$ do not affect the filamentation of electromagnetic waves \citep[][]{Sobacchi+2024a}, Eq.~\eqref{eq:envaux} is equivalent to
\begin{equation}
\label{eq:final5}
\frac{\rm i}{\omega_{\rm P}}\frac{\pa a}{\pa t} = -\frac{1}{2\omega_{\rm P}^2}\nabla^2a +\frac{1}{2} \delta\rho a \;.
\end{equation}
The evolution of the system is governed by Eqs.~\eqref{eq:final1}, \eqref{eq:final2}, \eqref{eq:final3}, \eqref{eq:final4}, \eqref{eq:final5}. When the plasma is unmagnetized (i.e.,~$\omega_{\rm L}=0$), one would recover the evolution model of \citet[][]{Sobacchi+2024a}.

\section{Filamentation instability}
\label{sec:filamentation}

Our final set of equations (i.e.,~Eqs.~\ref{eq:final1}, \ref{eq:final2}, \ref{eq:final3}, \ref{eq:final4}, \ref{eq:final5}) have an exact solution, which is $\delta\omega_{{\rm L}z}=\delta {\bm v}_\perp=\delta j_z=\delta \rho=0$ and $a=a_0$ (where $a_0$ is a constant, which we assumed to be real). In the following, we show that this solution is unstable, and the wave breaks into filaments. We study the evolution of a small perturbation of the wave intensity by defining $a=a_0 (1+\delta a)$. It is convenient to derive from Eq.~\eqref{eq:final5} two equations for $\delta a+\delta a^*$ and $\delta a-\delta a^*$, where $\delta a^*$ denotes the complex conjugate of $\delta a$. Assuming that $\delta a+\delta a^*$ and $\delta a-\delta a^*$ are proportional to $\exp[{\rm i}({\bm K}\cdot{\bm x}-\Omega t)]$, one finds
\begin{equation}
\label{eq:deltaa}
\frac{\Omega}{\omega_{\rm P}} \left(\delta a+\delta a^*\right) = \frac{K^2}{2\omega_{\rm P}^2}\left(\delta a-\delta a^*\right) \;, \qquad
\frac{\Omega}{\omega_{\rm P}} \left(\delta a-\delta a^*\right) = \frac{K^2}{2\omega_{\rm P}^2}\left(\delta a+\delta a^*\right) + \delta\rho \;.
\end{equation}
To determine the dispersion relation of the filamentation instability, one should use Eqs.~\eqref{eq:final1}-\eqref{eq:final4} to express $\delta\rho$ as a function of $\delta a+\delta a^*$. Taking into account that $|a|^2=a_0^2 (1+\delta a+\delta a^*)$ at leading order, one finds
\begin{equation}
\label{eq:deltarho}
\left(\Omega+K_z\right)^2 \delta\rho = \left[K_x^2\left(1+\frac{\omega_{\rm L}^2}{\omega_{\rm P}^2}\right) +\frac{K_y^2}{1-\frac{\omega_{\rm L}^2}{\gamma_0^2\left(\Omega+K_z\right)^2}}\right] \frac{a_0^2}{4\gamma_0^2}\left(\delta a+\delta a^*\right) \;.
\end{equation}
The dispersion relation can be determined from Eqs.~\eqref{eq:deltaa}-\eqref{eq:deltarho}. One finds
\begin{equation}
\label{eq:DRfil}
\left(\Omega+K_z\right)^2 \left(\frac{4\Omega^2}{K^2} - \frac{K^2}{\omega_{\rm P}^2}\right) = \frac{a_0^2}{2\gamma_0^2} \left[K_x^2\left(1+\frac{\omega_{\rm L}^2}{\omega_{\rm P}^2}\right) +\frac{K_y^2}{1-\frac{\omega_{\rm L}^2}{\gamma_0^2\left(\Omega+K_z\right)^2}}\right]  \;.
\end{equation}

To determine the growth rate of the filamentation instability, one should substitute $\Omega=-K_z+\Delta\Omega$ into Eq.~\eqref{eq:DRfil}, and consider long wavelengths in the longitudinal direction (i.e.,~$K_z\ll K_\perp$ and $K_z\ll \Delta\Omega$). First, we consider modes where the perturbation wave vector, $\bm{K}$, is parallel to the direction of the background magnetic field. This case corresponds to $K_y=0$. The maximal growth rate is given by
\begin{equation}
\label{eq:growthKx}
\left(\Delta\Omega\right)^2=-\frac{a_0^2\omega_{\rm P}^2}{2\gamma_0^2} \left(1+\frac{\omega_{\rm L}^2}{\omega_{\rm P}^2}\right) = -\frac{a_0^2\omega_{\rm P}^2}{2\gamma_0^2} \left(1+\sigma \right) \;,
\end{equation}
and it is achieved for wave numbers
\begin{equation}
\label{eq:maxKx}
K_x^2\gg \frac{a_0\omega_{\rm P}^2}{\gamma_0} \sqrt{1+\frac{\omega_{\rm L}^2}{\omega_{\rm P}^2}} = \frac{a_0\omega_{\rm P}^2}{\gamma_0} \sqrt{1+\sigma} \;.
\end{equation}
The growth rate, $|\Delta\Omega|$, which is given by Eq.~\eqref{eq:growthKx}, is larger by a factor $\sqrt{1+\sigma}$ with respect to our previous study of the filamentation instability in the non-relativistic limit $a_0\ll 1$ \citep{Sobacchi+2022}. The wave number, $K_x$, which is given by Eq.~\eqref{eq:maxKx}, is larger by a factor $(1+\sigma)^{1/4}$. The discrepancy is due to the fact that in our previous study we incorrectly neglected the secular evolution of the background magnetic field (our previous results are obtained by assuming $\delta\omega_{{\rm L}z}= 0$ in Eq.~\ref{eq:final4}).

In the following, we explain the reason why the growth rate of the filamentation instability is modified for large $\sigma$. The instability develops because the plasma density increases outside radiation filaments, and decreases inside the filaments. The resulting modulation of the plasma refractive index creates a converging lens that further enhances the radiation intensity inside the filaments, thus forming a positive feedback loop. When $\sigma<1$, density modulations arise because the transverse component of the ponderomotive force pushes the plasma outside radiation filaments (see Eq.~\ref{eq:final2}). When $\sigma>1$, another effect becomes dominant. The longitudinal component of the ponderomotive force pushes the plasma in the direction of wave propagation, and consequently the plasma is compressed \citep{GunnOstriker1971, Sobacchi+2024b}. When the perturbations of the wave intensity have a small amplitude, the longitudinal component of the ponderomotive force is nearly constant. Since filamentation distorts the background magnetic field lines (see Eq.~\ref{eq:final1}), there is an additional force acting on the plasma as a result of magnetic tension. Outside radiation filaments, the tension force has the same direction as the ponderomotive force (i.e.,~the tension force also pushes the plasma in the direction of propagation). Then, the density of the plasma increases as a result of magnetic tension. Instead, inside the filaments the tension force is antiparallel to the direction of propagation, and consequently the density decreases.

Now, we consider modes where $\bm{K}$ is perpendicular to the direction of the background magnetic field. This case corresponds to $K_x=0$. When $\gamma_0\Delta\Omega\gg\omega_{\rm L}$, the maximal growth rate is given by
\begin{equation}
\label{eq:growthKy}
\left(\Delta\Omega\right)^2=-\frac{a_0^2\omega_{\rm P}^2}{2\gamma_0^2} \;,
\end{equation}
and it is achieved for wave numbers
\begin{equation}
\label{eq:maxKy}
K_y^2\gg \frac{a_0\omega_{\rm P}^2}{\gamma_0} \;.
\end{equation}
The condition $\gamma_0\Delta\Omega\gg\omega_{\rm L}$ is satisfied for $\omega_{\rm L}/\omega_{\rm P}\ll a_0$, or equivalently $\sigma \ll a_0^2$. In this case, the wave packet is broken into roughly cylindrical filaments (the axis of the filaments is along the direction of wave propagation). Instead, when $\sigma\gg a_0^2$ modes with $K_x=0$ become stable. In this case, the wave packet is broken into slabs perpendicular to the direction of the background magnetic field. Eqs.~\eqref{eq:growthKy}-\eqref{eq:maxKy} are consistent with our previous study of the filamentation instability in the non-relativistic limit $a_0\ll 1$ \citep{Sobacchi+2022}. Fully kinetic simulations of relativistic quasi-perpendicular shocks show that the electromagnetic precursor is broken into roughly cylindrical filaments when the magnetization of the upstream flow is small, whereas it is broken into slabs when the magnetization is large \citep{Sironi+2021, Iwamoto+2024}. These results are consistent with our findings.

\citet[][]{Boyd+1987} study the filamentation instability in magnetized electron-proton plasmas in the non-relativistic limit $a_0\ll 1$. When $\bm{K}$ is parallel to the direction of the background magnetic field, the growth rate of the instability and the most unstable wave number increase when the strength of the background magnetic field increases. When $\bm{K}$ is perpendicular, the growth rate is suppressed when the strength of the background field increases. These results are consistent with our findings.

\section{Implications for relativistic shocks}
\label{sec:shock}

\begin{figure}
\label{fig:scenario}
\centering
\includegraphics[width=0.85\textwidth]{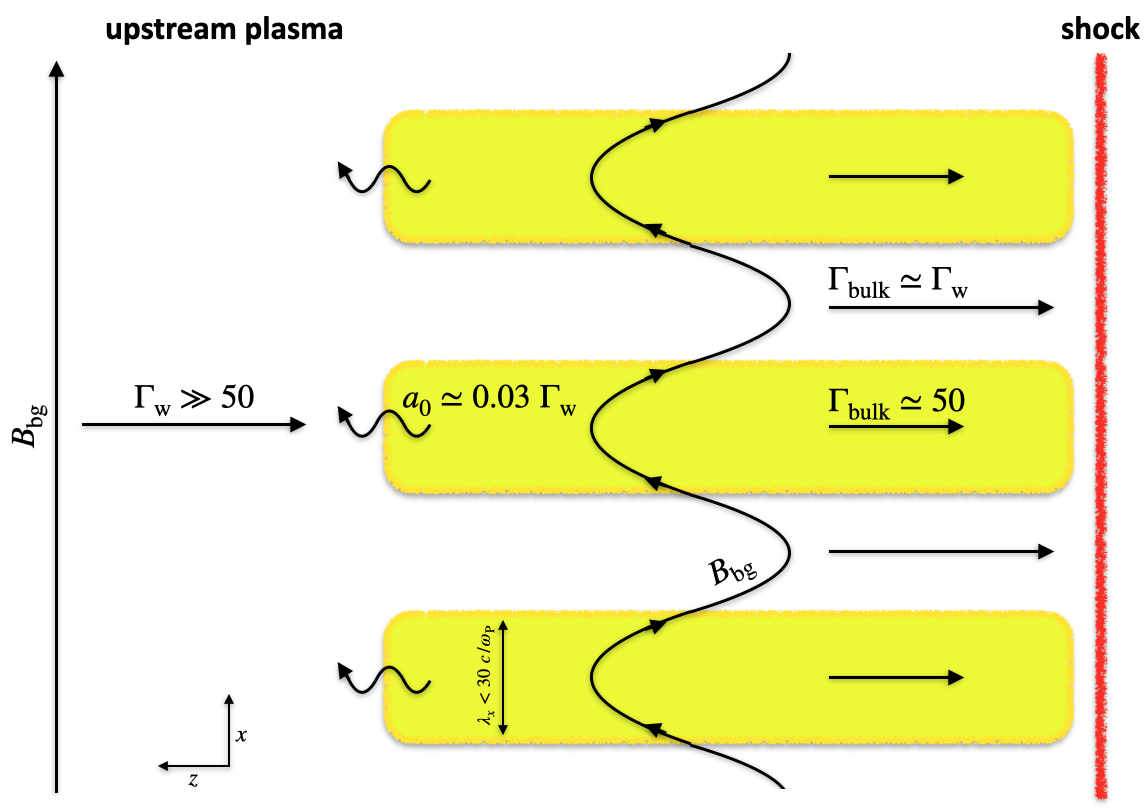}
\caption{Structure of relativistic quasi-perpendicular electron-positron shocks. For simplicity, we consider a magnetization $\sigma< 1$ (the general case is discussed in the text). The upstream plasma is threaded by a uniform background magnetic field. In the shock frame, the plasma has a large Lorentz factor, $\Gamma_{\rm w}\gg 50$. The shock produces an electromagnetic precursor with large strength parameter, $a_0\simeq 0.03\;\Gamma_{\rm w}$. The filamentation instability breaks the precursor into radiation filaments (indicated with the yellow regions), whose transverse scale is of the order of a few plasma skin depths. Inside radiation filaments, the bulk Lorentz factor of the plasma decreases to $\Gamma_{\rm bulk}\simeq 50$, whereas outside filaments one has $\Gamma_{\rm bulk}\simeq \Gamma_{\rm w}$. The relativistic shear flow distorts the background magnetic field lines, and produces a magnetic field component parallel to the shock normal that reverses across each filament (see Eq.~\ref{eq:final1}). The longitudinal scale of individual radiation filaments can be shorter than the total length of the precursor. In this case, the precursor would be broken into several incoherent filaments along the $z$ direction.}
\end{figure}

At the termination shock of the Crab pulsar wind, whose properties are summarized in Sect.~\ref{sec:precursor}, the electromagnetic precursor is almost certainly filamented. Consider a wave packet of length $\sim R_{\rm TS}$ in the pulsar frame, where the shock is stationary. The wave packet is moving with a Lorentz factor $\gamma_0=\omega_0/\omega_{\rm P}\simeq 3\;\Gamma_{\rm w}$ with respect to the wind (see Eq.~\ref{eq:omegaprec}). In the wave frame that moves with a Lorentz factor $\gamma_0$ with respect to the wind, the length of the wave packet is $\Delta R\sim \gamma_0 R_{\rm TS}/\Gamma_{\rm w}\sim 3\;R_{\rm TS}$. For perturbation wave vectors parallel to the background magnetic field, the growth rate of the filamentation instability in the wave frame can be calculated from Eq.~\eqref{eq:growthKx}, which gives $|\Delta\Omega| = a_0\omega_{\rm P}\sqrt{1+\sigma} /\gamma_0\sqrt{2}\simeq a_0\omega_{\rm P}\sqrt{1+\sigma} /3\sqrt{2}\Gamma_{\rm w}$. The instability develops when $|\Delta\Omega| \Delta R/c \gtrsim 10$, or equivalently $a_0\omega_{\rm P}R_{\rm TS}\sqrt{1+\sigma}/c\Gamma_{\rm w} \gtrsim 10\sqrt{2}$. Taking into account that $a_0\simeq 0.03\;\Gamma_{\rm w}/\sqrt{1+\sigma}$ (see Eq.~\ref{eq:aprec}), the latter condition is equivalent to
\begin{equation}
\label{eq:filcrab}
\frac{\omega_{\rm P}R_{\rm TS}}{c} \gtrsim 500 \;.
\end{equation}
Taking into account that $\omega_{\rm P}= 0.02\;\kappa_5\sqrt{1+\sigma}{\rm\; Hz}$, where $\kappa_5=\kappa/10^5$, and $R_{\rm TS}=4\times 10^{17}{\rm\; cm}$, Eq.~\eqref{eq:filcrab} implies $\kappa\gtrsim 200/\sqrt{1+\sigma}$. Then, the filamentation instability develops even for relatively small particle multiplicities.

As discussed in Sect.~\ref{sec:filamentation}, perturbation wave vectors perpendicular to the background magnetic field are unstable when $\sigma\lesssim a_0^2$. Taking into account that $a_0\simeq 2\times 10^4 (1+\sigma)^{-3/2} \kappa_5^{-1}$, the latter condition is equivalent to $\sigma\lesssim 140/\sqrt{\kappa_5}$, which could be satisfied in the equatorial region of the termination shock. Then, the precursor would be broken into roughly cylindrical filaments.

Filamentation of the electromagnetic precursor has important consequences for shock dynamics. After the instability reaches the non-linear stage, the radiation intensity could vanish outside the filaments. As we explain below, this would produce a strong velocity shear across each filament, because the bulk velocity of the upstream plasma depends on the local radiation intensity. The four-velocity and Lorentz factor of individual particles can be calculated in the test particle regime, neglecting the background magnetic field in the equation of motion \citep[][]{Sobacchi+2024b}. In this case, the solution of the equation of motion is well known \citep[][]{GunnOstriker1971}. In the proper frame of the wind, the $y$ and $z$ components of the four-velocity of the particles are $u_{y\pm}/c=\mp a$ and $u_{z\pm}/c=a^2/2$. The Lorentz factor is $\gamma_\pm=1+a^2/2$. The plus and minus signs respectively refer to positrons and electrons, and the vector potential of the electromagnetic wave is $amc^2/e$. In the shock frame, one has
\begin{equation}
\label{eq:testpart}
\frac{u_{y\pm}}{c} = \mp a \;, \qquad \frac{u_{z\pm}}{c} = -\Gamma_{\rm w}\beta_{\rm w} \left[1 - \frac{a^2}{2\Gamma_{\rm w}^2\beta_{\rm w}\left(1+\beta_{\rm w}\right)}\right] \;,\qquad \gamma_\pm = \Gamma_{\rm w}\left[1 + \frac{a^2}{2\Gamma_{\rm w}^2\left(1+\beta_{\rm w}\right)}\right] \;,
\end{equation}
where $\beta_{\rm w}=\sqrt{1-1/\Gamma_{\rm w}^2}$, and $a=a_0\cos\xi$. In the shock frame, the phase of the wave is $\xi= k_0(z-c t)/(1+\beta_{\rm w})\Gamma_{\rm w}$, where $k_0=\omega_0/c$ is the wave vector calculated in the proper frame of the wind (we assumed $\omega_0\gg\omega_{\rm P}$ based on Eq.~\ref{eq:omegaprec}). The bulk velocity of the flow along the shock normal is $\langle v_z\rangle =\langle u_{z\pm}/\gamma_\pm\rangle$, where the brackets indicate the average over many wavelengths. The bulk velocity can be conveniently approximated in the limit $\Gamma_{\rm w}\gg 1$ and $\Gamma_{\rm w}\gg a$ (the latter inequality is justified based on Eq.~\ref{eq:aprec}). Taking into account that $\langle a^2\rangle = a_0^2/2$, one finds $\langle v_z\rangle/c = -1 + (1+a_0^2/2)/2\Gamma_{\rm w}^2$. The bulk Lorentz factor, $\Gamma_{\rm bulk}=1/\sqrt{1-\langle v_z\rangle^2/c^2}$, is given by
\begin{equation}
\Gamma_{\rm bulk} = \frac{\Gamma_{\rm w}}{\sqrt{1+ a_0^2/2}} \;.
\end{equation}
Outside radiation filaments, where $a_0$ vanishes, one has $\Gamma_{\rm bulk}\simeq \Gamma_{\rm w}$. For $a_0\gg 1$, inside the filaments the bulk Lorentz factor significantly decreases and the flow becomes sheared (as discussed in Sect.~\ref{sec:precursor}, the condition $a_0\gg 1$ is almost certainly satisfied). Taking into account that $a_0\simeq 0.03\;\Gamma_{\rm w}/\sqrt{1+\sigma}$ (see Eq.~\ref{eq:aprec}), inside the filaments one has $\Gamma_{\rm bulk}\simeq 50\sqrt{1+\sigma}$. In the shock frame, the Lorentz factor of the particles, $\gamma_\pm$, is nearly unchanged when the plasma is illuminated by the electromagnetic precursor because $\Gamma_{\rm w}\gg a$ (see Eq.~\ref{eq:testpart}). However, as previously discussed by \citet[][]{Lyubarsky2018}, a significant fraction of the bulk kinetic energy of the wind can be transformed into internal energy.

The transverse scale of the filaments can be estimated from Eq.~\eqref{eq:maxKx}, which gives $\lambda_x\simeq \pi/K_x \lesssim (\pi c/\omega_{\rm P})(1+\sigma)^{-1/4}\sqrt{\omega_0/a_0\omega_{\rm P}}$. Taking into account that $\omega_0/a_0\omega_{\rm P}\simeq 100\sqrt{1+\sigma}$ (see Eqs.~\ref{eq:omegaprec} and \ref{eq:aprec}), one finds
\begin{equation}
\lambda_x \lesssim 30\;\frac{c}{\omega_{\rm P}} \simeq 5\times 10^{13} \left(1+\sigma\right)^{-1/2} \kappa_5^{-1} {\rm\; cm} \;.
\end{equation}
The transverse scale of the filaments can be larger than our estimate. Fully kinetic simulations show that in the non-relativistic limit $a_0\ll 1$ the filaments tend to merge when the instability reaches the non-linear stage \citep[][]{Iwamoto+2023}. We cannot capture this effect because our analysis focuses on the initial stages of the instability, when the perturbations of the wave intensity have a small amplitude.

The filamentation instability may be important for the acceleration of the non-thermal particles in relativistic quasi-perpendicular electron-positron shocks. The particles could be accelerated before entering the shock because the upstream flow is sheared, with significant variations of the bulk Lorentz factor on kinetic scales of the order of $\lambda_x$. The scenario is further enriched by the background magnetic field. Eq.~\eqref{eq:final1} shows that the filamentation instability distorts the field lines, and generates a magnetic field component parallel to the shock normal that reverses on the scale of the filaments, $\lambda_x$ (the source term on the right-hand side of Eq.~\eqref{eq:final1} is proportional to the gradient of the radiation intensity across filaments). Magnetic field reversals on such kinetic scales could trigger reconnection of the field lines, and consequently particles could be accelerated. In the wind frame the background magnetic field lines can be significantly distorted because in this frame the magnetic field of the precursor can be much larger than the background field (one has $B_0/B_{\rm bg}=a_0\omega_0/\omega_{\rm L}\simeq 3\times 10^{10} \sigma^{-1/2}(1+\sigma)^{-5/2}\kappa_5^{-2}$). Instead, in the shock frame the background magnetic field component parallel to the shock normal remains smaller than the perpendicular component.

The shock may exhibit a cyclic behavior. When the upstream plasma is cold, the precursor has a large strength parameter (given by Eq.~\ref{eq:aprec}). As discussed above, particles could be heated/accelerated before the plasma enters the shock. When the upstream plasma is filled with energetic particles, the efficiency of synchrotron maser emission is suppressed \citep[][]{BabulSironi2020}, and the strength parameter of the precursor decreases. The weakening of the precursor would suppress the acceleration of the particles, and the cycle would start over because the upstream plasma would be cold.

\section{Conclusions}
\label{sec:concl}

We studied the interaction of a strong electromagnetic precursor with the upstream plasma in relativistic quasi-perpendicular electron-positron shocks. The plasma is initially threaded by a uniform background magnetic field. We showed that the precursor can be broken into radiation filaments parallel to the shock normal. The transverse scale of the filaments is of the order of a few plasma skin depths. Since the bulk Lorentz factor of the plasma depends on the local radiation intensity, the filamentation instability produces a relativistic shear flow, where the bulk Lorentz factor changes significantly across each radiation filament. In the frame of the shock, outside radiation filaments the bulk Lorentz factor of the plasma is the same as in the absence of the precursor, whereas inside the filaments the bulk Lorentz factor is significantly reduced. The background magnetic field lines are distorted by the shear flow. Such distortion produces a magnetic field component parallel to the shock normal that reverses across each filament, a configuration that could trigger magnetic reconnection. Relativistic shear flow and magnetic reconnection may accelerate particles before the plasma enters the shock. Fully kinetic simulations are required to investigate this scenario.

\begin{acknowledgements}
We acknowledge insightful discussions with Elena Amato, Pasquale Blasi, Beno\^{i}t Cerutti, and Tsvi Piran. This work was supported by a Rita Levi Montalcini fellowship [E.S.], by the Simons Foundation Grant 00001470 to the Simons Collaboration on Extreme Electrodynamics of Compact Sources (SCEECS) [L.S.], by the DoE Early Career Award DE-SC0023015 [L.S.], by the Multimessenger Plasma Physics Center (MPPC) NSF Grant PHY2206609 [L.S.], by the NSF Grant AST-2307202 [L.S.], by the NASA ATP Grants 80NSSC24K1238 and 80NSSC24K1826 [L.S.], and by the JSPS KAKENHI Grant 22H00130 [M.I.].
\end{acknowledgements}

\bibliographystyle{aa}
\bibliography{2d}

\begin{thebibliography}{54}
\expandafter\ifx\csname natexlab\endcsname\relax\def\natexlab#1{#1}\fi

\bibitem[{{Amato} \& {Arons}(2006)}]{AmatoArons2006}
{Amato}, E. \& {Arons}, J. 2006, \apj, 653, 325

\bibitem[{{Babul} \& {Sironi}(2020)}]{BabulSironi2020}
{Babul}, A.-N. \& {Sironi}, L. 2020, \mnras, 499, 2884

\bibitem[{{Begelman} \& {Kirk}(1990)}]{BegelmanKirk1990}
{Begelman}, M.~C. \& {Kirk}, J.~G. 1990, \apj, 353, 66

\bibitem[{{Bourdier} \& {Fortin}(1979)}]{BourdierFortin1979}
{Bourdier}, A. \& {Fortin}, X. 1979, \pra, 20, 2154

\bibitem[{{Boyd} {et~al.}(1987){Boyd}, {Coutts}, \& {Marks}}]{Boyd+1987}
{Boyd}, T.~J.~M., {Coutts}, G.~A., \& {Marks}, D.~C. 1987, Physics of Fluids,
  30, 533

\bibitem[{{Bucciantini} {et~al.}(2011){Bucciantini}, {Arons}, \&
  {Amato}}]{Bucciantini+2011}
{Bucciantini}, N., {Arons}, J., \& {Amato}, E. 2011, \mnras, 410, 381

\bibitem[{{B{\"u}hler} \& {Blandford}(2014)}]{BuhlerBlandford2014}
{B{\"u}hler}, R. \& {Blandford}, R. 2014, Reports on Progress in Physics, 77,
  066901

\bibitem[{{Cerutti} \& {Giacinti}(2020)}]{CeruttiGiacinti2020}
{Cerutti}, B. \& {Giacinti}, G. 2020, \aap, 642, A123

\bibitem[{{Cerutti} {et~al.}(2020){Cerutti}, {Philippov}, \&
  {Dubus}}]{CeruttiPhilippov2020}
{Cerutti}, B., {Philippov}, A.~A., \& {Dubus}, G. 2020, \aap, 642, A204

\bibitem[{{Clemmow}(1974)}]{Clemmow1974}
{Clemmow}, P.~C. 1974, Journal of Plasma Physics, 12, 297

\bibitem[{{Del Zanna} {et~al.}(2004){Del Zanna}, {Amato}, \&
  {Bucciantini}}]{DelZanna+2004}
{Del Zanna}, L., {Amato}, E., \& {Bucciantini}, N. 2004, \aap, 421, 1063

\bibitem[{{Del Zanna} {et~al.}(2006){Del Zanna}, {Volpi}, {Amato}, \&
  {Bucciantini}}]{DelZanna+2006}
{Del Zanna}, L., {Volpi}, D., {Amato}, E., \& {Bucciantini}, N. 2006, \aap,
  453, 621

\bibitem[{{Drake} {et~al.}(1974){Drake}, {Kaw}, {Lee}, {Schmidt}, {Liu}, \&
  {Rosenbluth}}]{Drake+1974}
{Drake}, J.~F., {Kaw}, P.~K., {Lee}, Y.~C., {et~al.} 1974, Physics of Fluids,
  17, 778

\bibitem[{{Gaensler} \& {Slane}(2006)}]{GaenslerSlane2006}
{Gaensler}, B.~M. \& {Slane}, P.~O. 2006, \araa, 44, 17

\bibitem[{{Gallant} {et~al.}(1992){Gallant}, {Hoshino}, {Langdon}, {Arons}, \&
  {Max}}]{Gallant+1992}
{Gallant}, Y.~A., {Hoshino}, M., {Langdon}, A.~B., {Arons}, J., \& {Max}, C.~E.
  1992, \apj, 391, 73

\bibitem[{{Ghosh} {et~al.}(2022){Ghosh}, {Kagan}, {Keshet}, \&
  {Lyubarsky}}]{Ghosh+2022}
{Ghosh}, A., {Kagan}, D., {Keshet}, U., \& {Lyubarsky}, Y. 2022, \apj, 930, 106

\bibitem[{{Goldreich} \& {Julian}(1969)}]{GoldreichJulian1969}
{Goldreich}, P. \& {Julian}, W.~H. 1969, \apj, 157, 869

\bibitem[{{Gunn} \& {Ostriker}(1971)}]{GunnOstriker1971}
{Gunn}, J.~E. \& {Ostriker}, J.~P. 1971, \apj, 165, 523

\bibitem[{{Hester}(2008)}]{Hester2008}
{Hester}, J.~J. 2008, \araa, 46, 127

\bibitem[{{Hoshino} {et~al.}(1992){Hoshino}, {Arons}, {Gallant}, \&
  {Langdon}}]{Hoshino+1992}
{Hoshino}, M., {Arons}, J., {Gallant}, Y.~A., \& {Langdon}, A.~B. 1992, \apj,
  390, 454

\bibitem[{{Iwamoto} {et~al.}(2017){Iwamoto}, {Amano}, {Hoshino}, \&
  {Matsumoto}}]{Iwamoto+2017}
{Iwamoto}, M., {Amano}, T., {Hoshino}, M., \& {Matsumoto}, Y. 2017, \apj, 840,
  52

\bibitem[{{Iwamoto} {et~al.}(2018){Iwamoto}, {Amano}, {Hoshino}, \&
  {Matsumoto}}]{Iwamoto+2018}
{Iwamoto}, M., {Amano}, T., {Hoshino}, M., \& {Matsumoto}, Y. 2018, \apj, 858,
  93

\bibitem[{{Iwamoto} {et~al.}(2024){Iwamoto}, {Matsumoto}, {Amano}, {Matsukiyo},
  \& {Hoshino}}]{Iwamoto+2024}
{Iwamoto}, M., {Matsumoto}, Y., {Amano}, T., {Matsukiyo}, S., \& {Hoshino}, M.
  2024, \prl, 132, 035201

\bibitem[{{Iwamoto} {et~al.}(2023){Iwamoto}, {Sobacchi}, \&
  {Sironi}}]{Iwamoto+2023}
{Iwamoto}, M., {Sobacchi}, E., \& {Sironi}, L. 2023, \mnras, 522, 2133

\bibitem[{{Kargaltsev} {et~al.}(2015){Kargaltsev}, {Cerutti}, {Lyubarsky}, \&
  {Striani}}]{Kargaltsev+2015}
{Kargaltsev}, O., {Cerutti}, B., {Lyubarsky}, Y., \& {Striani}, E. 2015, \ssr,
  191, 391

\bibitem[{{Kaw} {et~al.}(1973){Kaw}, {Schmidt}, \& {Wilcox}}]{Kaw+1973}
{Kaw}, P., {Schmidt}, G., \& {Wilcox}, T. 1973, Physics of Fluids, 16, 1522

\bibitem[{{Kirk} \& {Skj{\ae}raasen}(2003)}]{KirkSk2003}
{Kirk}, J.~G. \& {Skj{\ae}raasen}, O. 2003, \apj, 591, 366

\bibitem[{{Komissarov} \& {Lyubarsky}(2003)}]{KomissarovLyubarsky2003}
{Komissarov}, S.~S. \& {Lyubarsky}, Y.~E. 2003, \mnras, 344, L93

\bibitem[{{Komissarov} \& {Lyubarsky}(2004)}]{KomissarovLyubarsky2004}
{Komissarov}, S.~S. \& {Lyubarsky}, Y.~E. 2004, \mnras, 349, 779

\bibitem[{{Langdon} {et~al.}(1988){Langdon}, {Arons}, \& {Max}}]{Langdon+1988}
{Langdon}, A.~B., {Arons}, J., \& {Max}, C.~E. 1988, \prl, 61, 779

\bibitem[{{Lyubarsky}(2018)}]{Lyubarsky2018}
{Lyubarsky}, Y. 2018, \mnras, 474, 1135

\bibitem[{{Lyubarsky}(2019)}]{Lyubarsky2019}
{Lyubarsky}, Y. 2019, \mnras, 490, 1474

\bibitem[{{Lyubarsky} \& {Kirk}(2001)}]{LyubarskyKirk2001}
{Lyubarsky}, Y. \& {Kirk}, J.~G. 2001, \apj, 547, 437

\bibitem[{{Lyubarsky} \& {Liverts}(2008)}]{LyubarskyLiverts2008}
{Lyubarsky}, Y. \& {Liverts}, M. 2008, \apj, 682, 1436

\bibitem[{{Lyubarsky}(2003)}]{Lyubarsky2003}
{Lyubarsky}, Y.~E. 2003, \mnras, 345, 153

\bibitem[{{Margalit} {et~al.}(2020){Margalit}, {Metzger}, \&
  {Sironi}}]{Margalit+2020}
{Margalit}, B., {Metzger}, B.~D., \& {Sironi}, L. 2020, \mnras, 494, 4627

\bibitem[{{Max} {et~al.}(1974){Max}, {Arons}, \& {Langdon}}]{Max+1974}
{Max}, C.~E., {Arons}, J., \& {Langdon}, A.~B. 1974, \prl, 33, 209

\bibitem[{{Olmi} {et~al.}(2015){Olmi}, {Del Zanna}, {Amato}, \&
  {Bucciantini}}]{Olmi+2015}
{Olmi}, B., {Del Zanna}, L., {Amato}, E., \& {Bucciantini}, N. 2015, \mnras,
  449, 3149

\bibitem[{{P{\'e}tri} \& {Lyubarsky}(2007)}]{PetriLyubarsky2007}
{P{\'e}tri}, J. \& {Lyubarsky}, Y. 2007, \aap, 473, 683

\bibitem[{{Plotnikov} \& {Sironi}(2019)}]{PlotnikovSironi2019}
{Plotnikov}, I. \& {Sironi}, L. 2019, \mnras, 485, 3816

\bibitem[{{Porth} {et~al.}(2014){Porth}, {Komissarov}, \&
  {Keppens}}]{Porth+2014}
{Porth}, O., {Komissarov}, S.~S., \& {Keppens}, R. 2014, \mnras, 438, 278

\bibitem[{{Reynolds} {et~al.}(2017){Reynolds}, {Pavlov}, {Kargaltsev},
  {Klingler}, {Renaud}, \& {Mereghetti}}]{Reynolds+2017}
{Reynolds}, S.~P., {Pavlov}, G.~G., {Kargaltsev}, O., {et~al.} 2017, \ssr, 207,
  175

\bibitem[{{Sironi} {et~al.}(2021){Sironi}, {Plotnikov}, {N{\"a}ttil{\"a}}, \&
  {Beloborodov}}]{Sironi+2021}
{Sironi}, L., {Plotnikov}, I., {N{\"a}ttil{\"a}}, J., \& {Beloborodov}, A.~M.
  2021, \prl, 127, 035101

\bibitem[{{Sironi} \& {Spitkovsky}(2009)}]{SironiSpitkovsky2009}
{Sironi}, L. \& {Spitkovsky}, A. 2009, \apj, 698, 1523

\bibitem[{{Sironi} \& {Spitkovsky}(2011)}]{SironiSpitkovsky2011}
{Sironi}, L. \& {Spitkovsky}, A. 2011, \apj, 741, 39

\bibitem[{{Sobacchi} {et~al.}(2024{\natexlab{a}}){Sobacchi}, {Iwamoto},
  {Sironi}, \& {Piran}}]{Sobacchi+2024b}
{Sobacchi}, E., {Iwamoto}, M., {Sironi}, L., \& {Piran}, T. 2024{\natexlab{a}},
  \aap, 690, A332

\bibitem[{{Sobacchi} {et~al.}(2024{\natexlab{b}}){Sobacchi}, {Iwamoto},
  {Sironi}, \& {Piran}}]{Sobacchi+2024a}
{Sobacchi}, E., {Iwamoto}, M., {Sironi}, L., \& {Piran}, T. 2024{\natexlab{b}},
  Physical Review Research, 6, 043213

\bibitem[{{Sobacchi} {et~al.}(2021){Sobacchi}, {Lyubarsky}, {Beloborodov}, \&
  {Sironi}}]{Sobacchi+2021}
{Sobacchi}, E., {Lyubarsky}, Y., {Beloborodov}, A.~M., \& {Sironi}, L. 2021,
  \mnras, 500, 272

\bibitem[{{Sobacchi} {et~al.}(2022){Sobacchi}, {Lyubarsky}, {Beloborodov}, \&
  {Sironi}}]{Sobacchi+2022}
{Sobacchi}, E., {Lyubarsky}, Y., {Beloborodov}, A.~M., \& {Sironi}, L. 2022,
  \mnras, 511, 4766

\bibitem[{{Sobacchi} {et~al.}(2023){Sobacchi}, {Lyubarsky}, {Beloborodov},
  {Sironi}, \& {Iwamoto}}]{Sobacchi+2023}
{Sobacchi}, E., {Lyubarsky}, Y., {Beloborodov}, A.~M., {Sironi}, L., \&
  {Iwamoto}, M. 2023, \apjl, 943, L21

\bibitem[{{Sprangle} {et~al.}(1990){Sprangle}, {Esarey}, \&
  {Ting}}]{Sprangle+1990}
{Sprangle}, P., {Esarey}, E., \& {Ting}, A. 1990, \prl, 64, 2011

\bibitem[{{Timokhin} \& {Harding}(2015)}]{TimokhinHarding2015}
{Timokhin}, A.~N. \& {Harding}, A.~K. 2015, \apj, 810, 144

\bibitem[{{Timokhin} \& {Harding}(2019)}]{TimokhinHarding2019}
{Timokhin}, A.~N. \& {Harding}, A.~K. 2019, \apj, 871, 12

\bibitem[{{Weisskopf} {et~al.}(2000){Weisskopf}, {Hester}, {Tennant}, {Elsner},
  {Schulz}, {Marshall}, {Karovska}, {Nichols}, {Swartz}, {Kolodziejczak}, \&
  {O'Dell}}]{Weisskopf+2000}
{Weisskopf}, M.~C., {Hester}, J.~J., {Tennant}, A.~F., {et~al.} 2000, \apjl,
  536, L81

\end{thebibliography}

\begin{appendix}

\section{Electrostatic potential}
\label{sec:charge}

In the following, we demonstrate that the electrostatic potential vanishes in the wave frame. Using Eq.~\eqref{eq:rho}, and taking into account that $|u_{{\rm f}y+}|=|u_{{\rm f}y-}|$ and $|u_{{\rm f}z+}|=|u_{{\rm f}z-}|$, the charge density can be presented as
\begin{equation}
\label{eq:chargedef}
\frac{\gamma_+n_+}{\gamma_0 n_0} - \frac{\gamma_-n_-}{\gamma_0 n_0} = \frac{\delta n_{0+}}{n_0} - \frac{\delta n_{0-}}{n_0} - \frac{u_0}{\gamma_0}\frac{\delta u_{0z+}}{\gamma_0} + \frac{u_0}{\gamma_0}\frac{\delta u_{0z-}}{\gamma_0} \;.
\end{equation}
The evolution of the charge density can be studied using Eq.~\eqref{eq:cont0aux}, which gives
\begin{equation}
\label{eq:charge}
\frac{\pa}{\pa t}\left( \frac{\delta n_{0+}}{n_0} - \frac{\delta n_{0-}}{n_0} - \frac{u_0}{\gamma_0}\frac{\delta u_{0z+}}{\gamma_0} + \frac{u_0}{\gamma_0}\frac{\delta u_{0z-}}{\gamma_0} \right) + \frac{\pa}{\pa z} \left( \frac{\delta u_{0z+}}{\gamma_0}-\frac{\delta u_{0z-}}{\gamma_0} -\frac{u_0}{\gamma_0}\frac{\delta n_{0+}}{n_0} + \frac{u_0}{\gamma_0}\frac{\delta n_{0-}}{n_0} \right) = - \nabla_\perp \cdot \left( \frac{\delta{\bm u}_{0\perp +}}{\gamma_0} - \frac{\delta{\bm u}_{0\perp -}}{\gamma_0} \right) \;.
\end{equation}
Eq.~\eqref{eq:charge} can be simplified as follows. One can use Eq.~\eqref{eq:deltauperp} to express $\delta{\bm u}_{0\perp +} - \delta{\bm u}_{0\perp -}$ as a function of ${\bm\psi}_{0\perp}$. Then, one can use the gauge condition, which is $\nabla_\perp\cdot {\bm\psi}_{0\perp} = -\pa\psi_{0z}/\pa z$, to eliminate ${\bm\psi}_{0\perp}$. Finally, one can use the $z$ component of Eq.~\eqref{eq:psi0} to eliminate $\psi_{0z}$. This procedure gives
\begin{equation}
\label{eq:nocharge}
\frac{\pa}{\pa t}\left( \frac{\delta n_{0+}}{n_0} - \frac{\delta n_{0-}}{n_0} - \frac{u_0}{\gamma_0}\frac{\delta u_{0z+}}{\gamma_0} + \frac{u_0}{\gamma_0}\frac{\delta u_{0z-}}{\gamma_0} \right) = 0 \;.
\end{equation}
From Eqs.~\eqref{eq:chargedef} and \eqref{eq:nocharge}, one finds $\pa (\gamma_+n_+-\gamma_-n_-) /\pa t =0$, which implies $\gamma_+n_+=\gamma_-n_-$. The electrostatic potential, $\phi$, is governed by Gauss's law, which is $\nabla^2\phi=-4\pi e (\gamma_+n_+-\gamma_-n_-)$. Since $\gamma_+n_+=\gamma_-n_-$, one finds $\phi=0$.

\section{Fluid velocity}
\label{sec:motion}

The equation that governs the evolution of $\delta{\bm u}_{0+}+\delta{\bm u}_{0-}$ can be derived from Eq.~\eqref{eq:u0}. One finds
\begin{align}
\nonumber
\frac{\pa}{\pa t} \left( \frac{\delta {\bm u}_{0+}}{\gamma_0} + \frac{\delta {\bm u}_{0-}}{\gamma_0} \right) & +\frac{u_0}{\gamma_0} {\bm e}_z \times \left[\nabla\times \left( \frac{\delta {\bm u}_{0+}}{\gamma_0} + \frac{\delta {\bm u}_{0-}}{\gamma_0} \right) \right] = \\
\label{eq:v0aux}
& = \nabla\left[ \frac{u_0}{\gamma_0}\left(\frac{\delta u_{0z+}}{\gamma_0} + \frac{\delta u_{0z+}}{\gamma_0}\right) - \frac{\left|a\right|^2}{2\gamma_0^2} \right] + \frac{1}{\gamma_0^2}\left(\frac{\delta u_{0z+}}{\gamma_0}-\frac{\delta u_{0z+}}{\gamma_0}\right)\omega_{\rm L}{\bm e}_y + \left( \frac{\delta {\bm u}_{0\perp+}}{\gamma_0} - \frac{\delta {\bm u}_{0\perp-}}{\gamma_0} \right) \times\omega_{\rm L}{\bm e}_x \;.
\end{align}
Taking into account that ${\bm e}_z\times (\nabla\times{\bm f})=\nabla_\perp f_z -\pa {\bm f}_\perp/\pa z$ for any function ${\bm f}$, and using Eq.~\eqref{eq:deltauperp} to express $\delta{\bm u}_{0\perp +}-\delta{\bm u}_{0\perp -}$ as a function of ${\bm\psi}_{0\perp}$, one sees that Eq.~\eqref{eq:v0aux} is equivalent to Eq.~\eqref{eq:v0}. The equation that governs the evolution of $\delta{\bm u}_{0+}-\delta{\bm u}_{0-}$ is
\begin{align}
\nonumber
\frac{\pa}{\pa t} & \left(\frac{\delta {\bm u}_{0+}}{\gamma_0} - \frac{\delta {\bm u}_{0-}}{\gamma_0} +\frac{2{\bm\psi}_0}{\gamma_0} \right) +\frac{u_0}{\gamma_0} {\bm e}_z \times \left[\nabla\times \left(\frac{\delta {\bm u}_{0+}}{\gamma_0} - \frac{\delta {\bm u}_{0-}}{\gamma_0}+\frac{2{\bm\psi}_0}{\gamma_0} \right) \right] + {\rm i}\frac{2\omega_{\rm L}}{\omega_{\rm P}}\left[ \frac{a}{2\gamma_0}\frac{\pa}{\pa y}\left(\frac{a^*}{2\gamma_0}\right) - \frac{a^*}{2\gamma_0}\frac{\pa}{\pa y}\left(\frac{a}{2\gamma_0}\right) \right] {\bm e}_z = \\
\label{eq:B0aux}
& = \frac{u_0}{\gamma_0}\nabla \left(\frac{\delta u_{0z+}}{\gamma_0} - \frac{\delta u_{0z-}}{\gamma_0} \right) + \left[\frac{\delta {\bm u}_{0\perp+}}{\gamma_0} + \frac{\delta {\bm u}_{0\perp-}}{\gamma_0} +\frac{1}{\gamma_0^2}\left(\frac{\delta u_{0z+}}{\gamma_0}+\frac{\delta u_{0z-}}{\gamma_0}\right){\bm e}_z +\frac{u_0}{\gamma_0}\frac{\left|a\right|^2}{2\gamma_0^2}{\bm e}_z \right]\times\omega_{\rm L}{\bm e}_x \;,
\end{align}
which is equivalent to Eq.~\eqref{eq:B0}.

\section{A useful identity}

To simplify the right-hand side of Eq.~\eqref{eq:v0}, one can use the identity demonstrated below. The gauge condition $\nabla\cdot{\bm\psi}_0=0$ implies $\pa^2\psi_{0y}/\pa y^2=-\pa^2\psi_{0x}/\pa x\pa y-\pa^2\psi_{0z}/\pa y\pa z$. Using the latter identity, and taking into account that $\delta \omega_{{\rm L}x}=\pa\psi_{0z}/\pa y-\pa\psi_{0y}/\pa z$ and $\delta \omega_{{\rm L}z}=\pa\psi_{0y}/\pa x-\pa\psi_{0x}/\pa y$, one can show that $\nabla^2\psi_{0y}=\pa\delta \omega_{{\rm L}z}/\pa x - \pa\delta \omega_{{\rm L}x}/\pa z$. Since $\pa\psi_{0y}/\pa t=-\delta\omega_{{\rm E}y}$, one finds
\begin{equation}
\nabla^2\psi_{0y}- \frac{\pa^2\psi_{0y}}{\pa t^2} = \frac{\pa\delta \omega_{{\rm L}z}}{\pa x} - \frac{\pa\delta \omega_{{\rm L}x}}{\pa z} +\frac{\pa \delta \omega_{{\rm E}y}}{\pa t} \;.
\end{equation}
Now, we can use Eq.~\eqref{eq:Ey} to eliminate $\delta\omega_{{\rm E}y}$, and then Eq.~\eqref{eq:Bxaux} to eliminate $\pa\delta\omega_{{\rm L}x}/\pa t$. This procedure gives
\begin{equation}
\label{eq:identity}
\nabla^2\psi_{0y} -\frac{\pa^2\psi_{0y}}{\pa t^2} = \frac{\pa\delta\omega_{{\rm L}z}}{\pa x} -\frac{u_0}{\gamma_0} \frac{\pa\delta\omega_{{\rm E}z}}{\pa y} -\frac{1}{\gamma_0^2} \frac{\pa\delta\omega_{{\rm L}x}}{\pa z} -\frac{\pa}{\pa t} \left( \frac{\left|a\right|^2}{4\gamma_0}\omega_{\rm L}+ \frac{\delta v_z}{\gamma_0}\omega_{\rm L} \right) -\frac{u_0}{\gamma_0}\frac{\pa}{\pa z} \left( \frac{\left|a\right|^2}{4\gamma_0}\omega_{\rm L}+ \frac{\delta v_z}{\gamma_0}\omega_{\rm L} \right)  \;.
\end{equation}

\section{Final set of equations}
\label{sec:finaleqs}

In the following, we derive Eqs.~\eqref{eq:final1}-\eqref{eq:final4}. Neglecting the time derivative of $|a|$, and taking into account that $u_0\simeq\gamma_0$, Eq.~\eqref{eq:divEfinal} can be approximated as
\begin{equation}
\label{eq:divEapp}
\frac{\pa^2}{\pa x^2}\left(\frac{\delta\omega_{{\rm E}z}}{\gamma_0\omega_{\rm L}}\right) + \frac{\pa^2}{\pa y^2}\left(\frac{\delta\omega_{{\rm E}z}}{\gamma_0\omega_{\rm L}}\right) + \frac{\pa^2}{\pa z^2}\left(\frac{\delta\omega_{{\rm E}z}}{\gamma_0\omega_{\rm L}}\right) - \frac{\pa^2}{\pa t\pa z}\left(\frac{\delta\omega_{{\rm E}z}}{\gamma_0\omega_{\rm L}}\right) = -\frac{\pa^2}{\pa t\pa y} \left(\frac{\delta v_z}{\gamma_0^2}\right)  \;.
\end{equation}
We introduce an additional approximation, and neglect terms of the order of $\delta v_z/\gamma_0^2$, as appropriate because $\gamma_0\gg 1$. In this approximation, Eq.~\eqref{eq:divEapp} gives $\delta\omega_{{\rm E}z}\simeq 0$. Substituting $\delta\omega_{{\rm E}z}\simeq 0$ into Eq.~\eqref{eq:Byaux}, one finds $\delta\omega_{{\rm L}y}\simeq 0$. Substituting $\delta\omega_{{\rm E}z}\simeq 0$ into Eq.~\eqref{eq:Bxaux}, and neglecting terms of the order of $\delta v_z/\gamma_0^2$, one finds
\begin{equation}
\label{eq:Bxapp}
\frac{\pa}{\pa t}\left(\frac{\delta\omega_{{\rm L}x}}{\gamma_0\omega_{\rm L}}\right) - \frac{\pa}{\pa z} \left( \frac{\delta\omega_{{\rm L}x}}{\gamma_0\omega_{\rm L}} - \frac{\left|a\right|^2}{4\gamma_0^2} \right) = 0 \;.
\end{equation}
Neglecting the time derivative of $|a|$, Eq.~\eqref{eq:Bxapp} gives $\delta\omega_{{\rm L}x}/\gamma_0\omega_{\rm L}\simeq |a|^2/4\gamma_0^2$.

Eq.~\eqref{eq:divBfinal} is equivalent to Eq.~\eqref{eq:final1} because terms of order $\delta v_z/\gamma_0^2$ can be neglected. Eq.~\eqref{eq:deltavperp} is clearly equivalent to Eq.~\eqref{eq:final2}. Taking into account that $\delta\omega_{{\rm E}z}\simeq 0$, Eq.~\eqref{eq:deltajz} can be presented as
\begin{equation}
\label{eq:jzapp}
\frac{\pa}{\pa t} \left(\frac{\delta j_z}{\omega_{\rm L}}\right) - \frac{\pa}{\pa z} \left(\frac{\delta j_z}{\omega_{\rm L}}\right) = - \delta v_y  - \frac{{\rm i}}{\omega_{\rm P}}\left[ \frac{a}{2\gamma_0}\frac{\pa}{\pa y}\left(\frac{a^*}{2\gamma_0}\right) - \frac{a^*}{2\gamma_0}\frac{\pa}{\pa y}\left(\frac{a}{2\gamma_0}\right) \right] \;.
\end{equation}
The second term on the right-hand side of Eq.~\eqref{eq:jzapp} can be neglected because the wave envelope, $a$, varies on spatial scales $\gg 1/\omega_{\rm P}$. In this limit, Eq.~\eqref{eq:jzapp} is equivalent to Eq.~\eqref{eq:final3}.

It is convenient to derive a new equation for the evolution of $\delta\tilde{\rho}$ by adding Eqs.~\eqref{eq:deltavzaux} and \eqref{eq:deltarhoaux}. Neglecting terms of the order of $\delta v_z/\gamma_0^2$, and taking into account that $\delta\omega_{{\rm E}z}\simeq 0$, one finds
\begin{equation}
\label{eq:rhoapp}
\frac{\pa}{\pa t}\left( \delta\tilde{\rho} + \frac{\left|a\right|^2}{4\gamma_0^2} - \frac{\omega_{\rm L}^2}{\omega_{\rm P}^2}\frac{\left|a\right|^2}{4\gamma_0^2} \right) - \frac{\pa}{\pa z}\left( \delta\tilde{\rho} - \frac{\left|a\right|^2}{4\gamma_0^2} -\frac{\omega_{\rm L}^2}{\omega_{\rm P}^2}\frac{3\left|a\right|^2}{4\gamma_0^2} \right) = -\nabla_\perp\cdot\delta {\bm v}_\perp +\frac{\omega_{\rm L}^2}{\omega_{\rm P}^2}\frac{\pa}{\pa x}\left(\frac{\delta\omega_{{\rm L}z}}{\gamma_0\omega_{\rm L}}\right) - \frac{\omega_{\rm L}^2}{\gamma_0^2\omega_{\rm P}^2}\frac{\pa}{\pa z}\left(\frac{\delta\omega_{{\rm L}x}}{\gamma_0\omega_{\rm L}}\right)  \;.
\end{equation}
Taking into account that $\delta\omega_{{\rm L}x}/\gamma_0\omega_{\rm L}\simeq |a|^2/4\gamma_0^2$, one sees that the last term on the right-hand side of Eq.~\eqref{eq:rhoapp} can be neglected, as it is of order $a_0^2/\gamma_0^4$. Then, Eq.~\eqref{eq:rhoapp} can be presented as
\begin{equation}
\label{eq:rhoappaux}
\frac{\pa}{\pa t}\left( \delta\rho + \frac{\left|a\right|^2}{2\gamma_0^2} + \frac{\omega_{\rm L}^2}{\omega_{\rm P}^2}\frac{\left|a\right|^2}{2\gamma_0^2} \right) - \frac{\pa\delta\rho}{\pa z} = -\nabla_\perp\cdot\delta {\bm v}_\perp +\frac{\omega_{\rm L}^2}{\omega_{\rm P}^2}\frac{\pa}{\pa x}\left(\frac{\delta\omega_{{\rm L}z}}{\gamma_0\omega_{\rm L}}\right) \;,
\end{equation}
where $\delta\rho$ is defined by Eq.~\eqref{eq:rhotildedef}. Since the time derivative of $|a|$ can be neglected, Eq.~\eqref{eq:rhoappaux} is equivalent to Eq.~\eqref{eq:final4}.

\end{appendix}

\end{document}